\documentclass[reprint,amsmath,amssymb,prl,aps,superscriptaddress]{revtex4-2}
\usepackage{amssymb}
\usepackage{amsmath,bm}
\usepackage{graphicx,color}
\usepackage{xcolor}
\usepackage[caption=false]{subfig}
\usepackage[normalem]{ulem} 
\usepackage{graphicx}  
\usepackage{tikz}
\usepackage[percent]{overpic}
\usepackage{fancybox} 
\usepackage{hyperref}
\usepackage{booktabs}
\usepackage{graphicx}
\usepackage{fancybox} % for \ovalbox
\usepackage{cancel}
\usepackage{xr}
\newcommand{\figref}[1]{Fig.~\ref{#1}}

\newcommand{\brk}[1]{\left(#1\right)}

\newcommand{\nabar}{\Bar{\nabla}}
\newcommand{\abar}{\Bar{a}}

\newcommand{\m}{{m}}
\newcommand{\s}{{s}}

\newcommand{\A}{\mathcal{A}}

\newcommand{\appref}[1]{%
Sec.~\ifcsname smsec@#1\endcsname
\csname smsec@#1\endcsname
\else ??\fi\ of the SM~\cite{SM}%
}

\expandafter\def\csname smsec@numerical_methods\endcsname{I}
\expandafter\def\csname smsec@full_eq\endcsname{II}
\expandafter\def\csname smsec@b_t\endcsname{III}
\expandafter\def\csname smsec@clamped_b_c\endcsname{III}
\expandafter\def\csname smsec@isometry\endcsname{IV}
\expandafter\def\csname smsec@compressed_ribbon\endcsname{V}
\expandafter\def\csname smsec@app:Euler_elastica\endcsname{V}
\expandafter\def\csname smsec@twisted_derivation\endcsname{VI}
\expandafter\def\csname smsec@open_disc\endcsname{VII}

\renewcommand{\div}{\mathrm{div}_{\abar,a}}

\begin{document}

\title{Curvature Potential Formulation for Thin Elastic Sheets}

\author{Yael Cohen}
 \affiliation{Racah Institute of Physics, The Hebrew University of Jerusalem, Jerusalem, Israel}
 
 \author{Animesh Pandey}
 \affiliation{Engineering Mechanics Unit, Jawaharlal Nehru Center for Advanced Scientific Research, Bengaluru, India}

 \author{Yafei Zhang}
 \affiliation{Racah Institute of Physics, The Hebrew University of Jerusalem, Jerusalem, Israel}

\author{Cy Maor}
 \affiliation{Einstein Institute of Mathematics, The Hebrew University of Jerusalem, Jerusalem, Israel}
 
\author{Michael Moshe}
 \email{michael.moshe@mail.huji.ac.il} \affiliation{Racah Institute of Physics, The Hebrew University of Jerusalem, Jerusalem, Israel}

\begin{abstract}
 {Thin sheets, from graphene to biological membranes, exhibit wrinkling, folding, and thermal deformations driven by geometric nonlinearities. They are treated using weakly nonlinear theories such as the Föppl-von Kármán equations, whose reliance on small slopes and a single-valued height field excludes large-slope and non-graph configurations even at small strain. We introduce a geometric reformulation in terms of scalar stress and curvature potentials that preserves the FvK structure while removing the graph restriction. We show that intrinsic symmetries of these potentials, as opposed to symmetries of the configuration,  can persist when none is apparent in the embedding, enabling analytical solutions of boundary-value problems inaccessible to the conventional height-based formulation. Finite-element comparisons validate the predictions under strong confinement. Together, the extended validity and tractability open analytical routes to fluctuating membranes, ribbons, and architected sheets.}
\end{abstract}
\maketitle
\textbf{\textit{Introduction.}}
The physics of thin elastic sheets is central to a broad range of condensed-matter systems~\cite{van2024soft, nelson2002defects, chaikin1995principles}, 
from graphene, whose electronic properties are strain-dependent~\cite{levy2010strain,zhu2015programmable}, to biological membranes such as tissues and plant leaves ~\cite{sharon2004leaves,zhang2025geometrically,armon2011geometry, murisic2015discrete,dervaux2008morphogenesis}. 
Yet the elasticity of thin sheets is a geometrically nonlinear field theory, making analytical treatment challenging \cite{EFRATI2009762, landau2012theory}.
A common strategy for analytical progress is to use controlled approximations, representing the surface by in-plane displacements $\mathbf{d}$ and a height function $f$ describing out-of-plane deflection relative to a reference surface, together with a scaling relation between deformation modes.
In the weakly nonlinear regime, $|\nabla \mathbf{d}| \sim |\nabla f|^2 \ll 1$, this leads to the classical nonlinear Föppl–von Kármán (FvK) theory~\cite{landau2012theory, lewicka2011foppl,ciarletta2022foppl}. 

\begin{figure}
    \centering
    \includegraphics[
        width=\columnwidth]{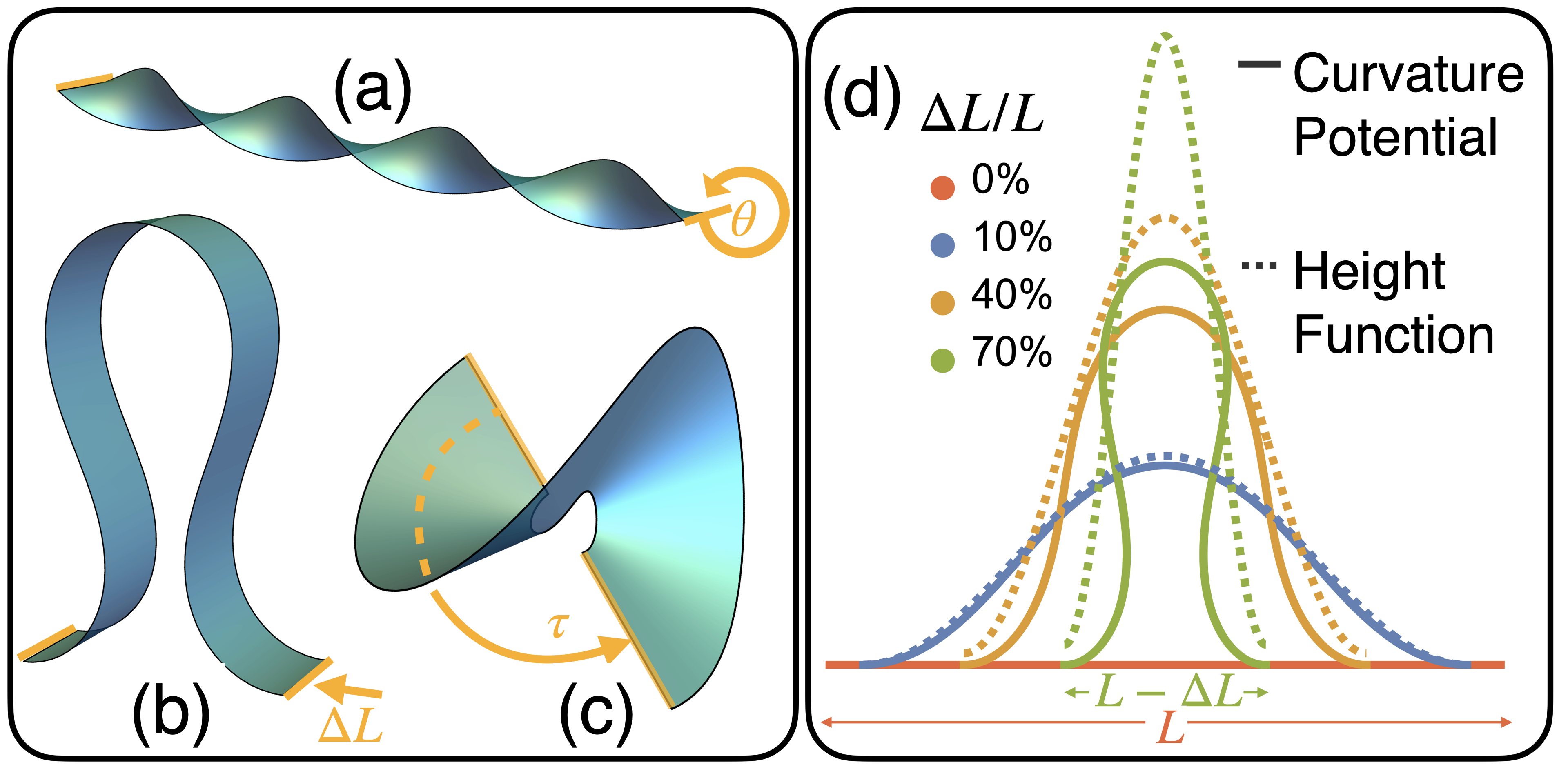}
    \caption{Left: Curvature-potential based analytical solutions for deformations that cannot be represented by a height function: (a) twisted ribbon  {with $\theta=4\pi$} , (b) compressed ribbon  {with $\Delta L/L=0.7$}, and (c) sliced annulus  {with $\tau=\pi$}. Right (d):
    Compressed-ribbon configurations from the curvature-potential formulation (solid) versus FvK theory (dashed).}
    \label{fig:motivation}
\end{figure}

This approach has been highly successful, underpinning results ranging from renormalization of elastic moduli in thermally fluctuating membranes~\cite{nelson1989statistical,PACHECOSANJUAN2019105154,kovsmrlj2016response,le2018anomalous}, to defect mechanics in crystalline monolayers~\cite{seung1988defects,zhang2014defects}, and extreme phenomena such as wrinkling and crumpling in slender solids~\cite{guo2025localized,leembruggen2023computational,chopin2015roadmap,cerda1998conical,cerda2003geometry,liang2009shape}.
Despite these achievements, it has inherent limitations:  {in addition to} the small-strain assumption, it is restricted to small slopes and to configurations representable as a single-valued height function $f$.
Many physically relevant systems, from kirigami architectures~\cite{sadik2021local} to graphene monolayers~\cite{zhang2014defects}, have low elastic energy yet violate these conditions. This is illustrated in Fig.~\ref{fig:motivation}, where prototypical deformation modes cannot be represented by a single-valued height function  {even at small strains}, so analysis typically requires returning to the full configuration-based formulation~\cite{kovsmrlj2016response}.  {A further limitation is that analytical solutions are largely restricted to highly symmetric configurations. Extending both the analytical tractability and validity to multi-valued configurations is therefore an important step toward a general framework for thin-sheet elasticity.}

In this Letter, we introduce a fully intrinsic geometric formulation of thin-sheet elasticity that applies seamlessly to configurations with large slopes or multi-valued height functions, as illustrated in \figref{fig:motivation}(a–c). 
 {While intrinsic geometric elasticity naturally describes growing and frustrated sheets, it is often analytically cumbersome. The curvature–potential approach, which is the topic of this work, retains the intrinsic geometric content while enabling analytical insight through scalar potentials. A major advantage is that} 
it preserves the mathematical structure of existing methods and provides a unified reinterpretation of previously studied problems. 
 {Furthermore, because the fundamental variables are intrinsic elastic fields rather than the embedding itself, the formulation admits analytical simplification in problems whose configurations exhibit no obvious geometric symmetry but whose elastic fields remain intrinsically symmetric. }
Crucially, it extends these analyses to complex geometries, boundary conditions, and geometric frustration beyond the reach of standard weakly nonlinear theories. 

{At the heart of our approach are two scalar fields: the Airy stress potential $\chi$ (SP), generating the in-plane stress tensor, and a curvature potential $\psi$ (CP), generating the surface curvature tensor. This representation extends the kinematic description to multi-valued embeddings and large rotations. 
In regimes where stretching is small but cannot be neglected (FvK), the resulting equilibrium equations expressed in terms of $\chi$ and $\psi$ retain the form of the established FvK theory, but with a CP interpretation rather than a height function.}

This distinction is illustrated in \figref{fig:motivation}(d), which shows two interpretations of the \emph{same} elastic-ribbon solution: the CP interpretation (solid line) and the FvK interpretation (dashed). The analytic solutions for twisting and bending in \figref{fig:motivation}(a,c) have no counterparts within the FvK  {interpretation}.

In what follows, we present the geometric formulation in terms of stress and curvature potentials and apply it to geometries beyond current frameworks. 
To validate the approach, we compare the analytical results with finite-element simulations of the full problem,  {using a numerical method developed and validated in earlier work~\cite{armon2014shape, moshe2019nonlinear}}, and find excellent agreement. 
 {These results establish the curvature–potential formulation as an analytic framework for the mechanics and geometry of thin-sheet elasticity.}

\textbf{\textit{The Theoretical Framework.}}
We start from a kinematic description of the configuration $\boldsymbol{\phi}(u,v) = \{X(u,v), Y(u,v), Z(u,v)\}$.
In geometric elasticity, the metric $a_{\alpha\beta}=\partial_\alpha \boldsymbol{\phi} \cdot \partial_\beta \boldsymbol{\phi}$ 
encodes distances between nearby points on the surface, while the curvature tensor $b_{\alpha\beta} = -\partial_\alpha \mathbf{n} \cdot \partial_\beta \boldsymbol{\phi}$ encodes the orientation of the surface normal $\mathbf{n}$. 
The elastic energy penalizes deviations from reference fields $(\bar a,\bar b)$, yielding the standard stretching–bending energy functional \footnote{The unusual factor of $1/2\sqrt{3}$ in the definition of $\epsilon_\mathrm{bn}$ replaces the factor of $1/12$ in the bending energy commonly arising from thickness integration through dimension reduction approaches, see \cite{EFRATI2009762}.}
\begin{eqnarray}\label{energy functional}
E &=& \intop \frac{1}{2}\left[  h \A^{\alpha\beta\gamma\delta} \epsilon^\mathrm{st}_{\alpha\beta} \epsilon^\mathrm{st}_{\gamma\delta} +  h^3 \A^{\alpha\beta\gamma\delta} \epsilon^\mathrm{bn}_{\alpha\beta} \epsilon^\mathrm{bn}_{\gamma\delta} \right] dS \;,\notag\\
\epsilon^\mathrm{st} &=& \tfrac{1}{2}(a - \Bar{a})\;, \quad \epsilon^\mathrm{bn} = \tfrac{1}{2\sqrt{3}}(b - \Bar{b})\;,
\end{eqnarray}
 {with $\A$ the elastic tensor and $h$ is the thickness of the sheet.}
Although the reference fields $\bar{a}$ and $\bar{b}$ are independent, the actual fields 
$a$ and $b$ arise from an embedding $\boldsymbol{\phi}$. They are therefore constrained by the Gauss and 
Mainardi-Peterson-Codazzi (MPC) compatibility conditions,
\begin{eqnarray}
    K_G(a)= \det(a^{-1} b) \;, \quad 
    \nabla \times b = 0
    \;, \label{eq:nb=nb}
\end{eqnarray} 
which ensure that $a$ and $b$ are locally realizable by a surface in $\mathbb{R}^3$. 
The first relation is Gauss' \emph{theorema egregium}, equating the intrinsic Gaussian curvature to its extrinsic form via the shape operator $s = a^{-1} b$. The second condition expresses the integrability of the normal field.

The equilibrium equations follow from minimizing  {the energy}~\eqref{energy functional} with respect to the embedding $\boldsymbol{\phi}(u,v)$, 
yielding the system~\cite{EFRATI2009762}
\begin{eqnarray}
\div \, \Sigma+ \div\, \m \, \s = 0 , \quad
{\text{div}}_{\abar} \left(\div\,\m \right)-\Sigma \, b = 0,\label{efis_eq2}
\end{eqnarray}
where $\Sigma=\sigma+\m\,s$ is an effective stress combining stretching and bending contributions. Here the stress $\sigma$ and bending moment $\m$ are linear in the strains $\epsilon^{\mathrm{st}}$ and $\epsilon^{\mathrm{bn}}$, and $\mathrm{div}_{\abar}$ denotes the covariant divergence with respect to $\abar$.
 {The primary fields in this formulation are $a$ and $b$, and the equilibrium equations Eq.~\ref{efis_eq2} are closed by the compatibility conditions Eq.~\ref{eq:nb=nb}.}

A key difficulty is that the differential operator $\div$ depends on the unknown metric $a$ (and on $\abar$), making the system nonlinear and implicit~\cite{moshe2014plane} (see \appref{full_eq}).
 {Nevertheless, geometric elasticity has yielded important analytic insights in a variety of problems such as morphogenesis and self-assembly~\cite{sharon2004leaves,audoly2005cracks,armon2011geometry,davidovitch2011prototypical,aharoni2012emergence,grason2016perspective,li2021chemically,wang2025geometrically}. This motivates seeking formulations that retain its geometric content while improving analytical tractability.}

The central idea underlying our approach is to enforce the compatibility equations directly, without introducing an explicit embedding $\boldsymbol{\phi}$ or a height function.
This is achieved by introducing a scalar CP that generates curvature fields automatically consistent with MPC compatibility. 
This construction is valid in the small-strain regime, regardless of whether the configuration admits a height-function representation or involves large slopes.
While the framework applies more broadly, for concreteness and clarity we restrict here to cases
with $\bar{a}$ close to Euclidean and $\bar{b}=0$.

We first note that for  {$\bar{a}$ close to Euclidean, to first order in small strains, the MPC conditions reduce to $\bar{\nabla}\times b = 0$. Thus, satisfying MPC is equivalent to}
\begin{eqnarray}\label{b_from_psi}
b_{\alpha\beta}&=&\nabar_{\alpha\beta}\psi\;.
\end{eqnarray}
Within this approximation, the divergence operators in~\eqref{efis_eq2} simplify to $\mathrm{div}_{\abar}$, and for small bending strains the in-plane equilibrium equation becomes $\mathrm{div}_{\abar}\,\Sigma=0$. 
This condition is satisfied identically by introducing a generalized Airy stress potential,
$\Sigma=\bar{\nabla}\times\bar{\nabla}\times\chi$~\cite{moshe2014plane}. 
The remaining constraints are the Gauss compatibility relation~\eqref{eq:nb=nb} and the bending-moment equation~\eqref{efis_eq2}, whose  {approximate} form depends on the scaling between stretching and bending strains $\epsilon_{\mathrm{st}}$ and $\epsilon_{\mathrm{bn}}$.

\begin{figure*}
    \centering
    \includegraphics[width=0.75\linewidth]{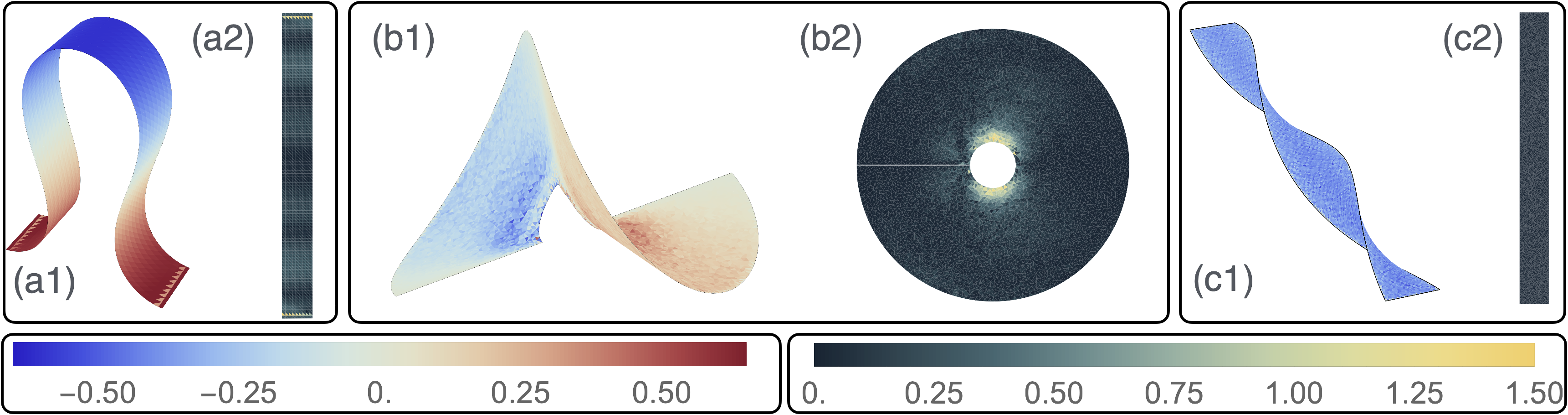}
    \caption{Equilibrium configurations from finite-element simulations and the corresponding curvature fields compared with predictions by the CP formulation, shown in the Lagrangian frame.
(a1) Compressed ribbon ($L=10$, $W=1$, $h=0.1$)  {with $\Delta L/L=0.7$}. (b1) Bent sliced annulus ($R_{\mathrm{in}}=1$, $R_{\mathrm{out}}=5$, $h=0.05$)
 {with an opening angle of $\tau=\pi$}. (c1) Twisted ribbon ($L=10$, $W=1$, $h=0.01$)  {with total twist of $\theta=2\pi$}.
Panels (a1) and (b1) are colored by mean curvature, and panel (c1) by Gaussian curvature. Relative deviations are shown in (a2) and (b2), $(H_{\mathrm{FE}}-H_{\mathrm{CP}})/\max(H_{\mathrm{CP}})$, and in (c2), $1-K_{\mathrm{FE}}/K_{\mathrm{CP}}$.}
    \label{fig:all_sims}
\end{figure*}

 {
In the FvK regime, relevant to many soft-matter systems, both stretching and bending are small but not negligible, $\epsilon_{\mathrm{st}} \ll 1$ and $h\,\epsilon_{\mathrm{bn}} \ll 1$. 
These conditions are satisfied when the sheet is thin compared with all in-plane length scales and the characteristic curvature $\kappa$ is small, so that $h\kappa \ll 1$.
}

The equilibrium equations then take the form
\begin{eqnarray}
 Y^{-1}\bar{\Delta} \bar{\Delta} \chi  &=& \bar{K}_G-\det(\bar{a}^{-1} b)\label{eq:chi}
     \\
B \bar{\Delta}\bar{\Delta} \psi&=& \sigma^{\alpha\beta} b_{\alpha\beta}\;. \label{eq:psi}
\end{eqnarray}
 {where $\bar{\Delta} = \bar{a}^{\alpha\beta}\bar{\nabla}_{\alpha\beta} $ is the covariant Laplacian (Laplace-Beltrami operator), }
{$Y$ and $\nu$ are the 2D Young's modulus and Poisson's ratio,
% $h$ is the thickness of the sheet, 
and the bending modulus $B={Y h^2}/{[12\brk{1-\nu^2}]}$.}
These equations coincide in form with the covariant FvK equations, with the crucial distinction that $\psi$ is a curvature potential rather than a configurational height function. 
In particular, this formulation permits large slopes, rotations, and deflections
\footnote{Other scaling assumptions modify these equations. For example, in the infinitely thin limit where $a=\abar$, the stress in Eq.~\eqref{eq:psi} plays the role of a Lagrange multiplier enforcing isometry (see \appref{isometry}).}.
 {It is worth noting that when the surface admits a graph representation $z=f(u,v)$, the small-slope approximation identifies the deflection field $f$ with the CP $\psi$}.
 {Given a solution for the potentials, the curvature tensor is obtained directly from $\psi$ through Eq.~\eqref{b_from_psi}, while the metric field $a$  {follows from the stress-strain relation, with the stress} $\sigma \approx \nabar\times\nabar\times\chi$. 
Since stresses and moments derive from $\chi$ and $\psi$, this formulation is natural for force- and torque-controlled boundaries.
 {Free boundaries correspond to vanishing normal stress and torque, $\Sigma\cdot n = 0$, $n \cdot  m\cdot n = 0$ and $\nabla m\cdot n=0$ \cite{EFRATI2009762} (see \appref{b_t}).}
Thus, solving the equilibrium problem for a sheet with non-Euclidean $\abar$ reduces to solving Eqs.~\eqref{eq:chi}–\eqref{eq:psi} subject to prescribed traction forces and torques.
Displacement-controlled boundary conditions can be imposed  {on $\phi$} upon reconstructing the configuration from the potentials.}

 {However, this reconstruction proceeds in a manner fundamentally different from the classical height-based approach.}
The equations determining the configuration from the fundamental forms are the classical Weingarten equations~\cite{DoCarmo1976}: Defining the tangent vectors $\mathbf{t}_\alpha = \partial_\alpha \boldsymbol{\phi}$ and the unit normal $\mathbf{n}$, this frame satisfies
\begin{eqnarray}
    \partial_\alpha\mathbf{n}&=& -b^\mu_{~\alpha}~\mathbf{t}_\mu \,, \quad 
   \partial_{\alpha} \mathbf{t}_\beta=\Gamma^\mu_{\alpha\beta}~\mathbf{t}_\mu+b_{\alpha\beta}\mathbf{n}
   \label{eq: Weingarten}
\end{eqnarray}
where $\Gamma$ is the Levi-Civita connection associated with $a$~\cite{DoCarmo1976}. Solving this linear system yields the frame $\{\mathbf{t}_1,\mathbf{t}_2,\mathbf{n}\}$, from which the surface embedding $\boldsymbol{\phi}(u,v)$ is recovered by integration.
Thus, once the potentials $\chi$ and $\psi$ are known, the fundamental forms $a$ and $b$, geometrically compatible by construction, are recovered, and the configuration follows from integrating Eq.~\ref{eq: Weingarten}.
 {
If traction forces and torques are prescribed, the reconstructed configuration is determined only up to rigid translations and rotations (i.e., the position and orientation of the seed point are arbitrary). This gauge freedom is removed once boundary conditions are imposed directly on the reconstructed configuration.
}

\textbf{\textit{The Curvature Potential in Action.}}
We  {first } illustrate the CP framework through three prototypical systems: a clamped compressed ribbon, a twisted ribbon, and a sliced annulus,  {demonstrating how the CP formulation extends the classical FvK theory to multi-valued configurations. We then consider mixed loading conditions which demonstrate a second advantage: although the embedding exhibits no obvious symmetry, the stress and curvature potentials are intrinsically symmetric, thereby extending the class of problems that admit analytical treatment. } While the framework applies to non-Euclidean plates, here we focus on boundary-condition effects in systems with Euclidean reference metric, leaving more general cases with non-Euclidean $\bar{a}$ for future work, as discussed in the Summary.

To benchmark the theory, we compare our predictions with finite-element simulations of the elastic energy~\eqref{energy functional}. 
The sheet is discretized into a triangulated mesh whose elements carry target metric, target curvature, thickness, and elastic moduli.  
For each configuration we compute the resulting metric and curvature, evaluate the elastic energy, and relax the shape subject to the prescribed boundary conditions. 
To validate the solutions, in each example we verify that bending remains small, $h\kappa \ll 1$, with $\kappa$ the characteristic curvature and $h$ the thickness, and quantify the discrepancy between analytical and numerical results via a relative energy deviation $\varepsilon$  (see \appref{numerical_methods}).
These simulations provide a quantitative reference for strongly nonlinear regimes beyond the reach of classical analytical approaches \cite{armon2014shape, moshe2019nonlinear}.

\paragraph{Compressed ribbon.}
A long, narrow ribbon  {of length $L$ compressed between clamped ends to a length $\ell = L - \Delta L$ provides} the classical Euler-buckling scenario, relevant to thermalized graphene sheets, polymer ribbons and more \cite{kovsmrlj2016response,yu2019bifurcations}.  
While the FvK equations describe the buckling threshold and the weakly 
post-buckled regime, they cease to apply under strong compression where slopes become large and eventually no longer representable as a height graph~\cite{AudolyPomeau2010-Ch8}.
 {
Assuming stress and curvature that can vary only along the ribbon we find the solution
\begin{equation}
\chi(u,v)=-B \lambda^2 v^{2}/{2},\quad
\psi(u,v)=A\cos(\lambda u),
\end{equation}
with $A$ and $\lambda$ determined by boundary conditions. The characteristic curvature in this solution is $\kappa = A \lambda^2$.} 

 {Upon interpreting $\psi$ as CP, and integrating \eqref{eq: Weingarten} to reconstruct the configuration, we obtain}
$\hat{n}(u) = \left(\sin(A\lambda\sin(\lambda u)),0,\cos(A\lambda\sin(\lambda u))\right)$.
For clamped ends we impose $\hat{n}(\pm L/2) = (0,0,1)$ which yields $\lambda=\frac{2\pi}{L}n$.
For the lowest mode $n=1$, the condition $\ell = L - \Delta L$  results with an equation for $A$, which satisfies $J_0(2\pi A/L) = 1-\Delta L/L$ ( {with $J_0$ the Bessel function of the first kind, } see \appref{clamped_b_c}).

In \figref{fig:motivation}(d) we show the profiles obtained when $\psi$ is treated as a CP and integrated through the Weingarten equations (solid line), versus a FvK height function (dashed line).
While the classical approach typically fails at strains of order $10\%$, our approach remains accurate even at strains as high as $70\%$ (i.e., $\ell=0.3L$),
 {with $h  \kappa \sim 0.1$, and $\varepsilon \approx 10^{-2}$.}

A detailed comparison with full finite-element solution is shown in \figref{fig:all_sims}(a): panel (a1) shows the configuration colored by mean curvature, 
and \figref{fig:all_sims}(a2) displays the  {small} difference between analytical and numerical mean curvature across the ribbon in the Lagrangian frame.
 {Note that } in the isometric limit where stretching vanishes, the solution reduces to Euler elastica. 
In \appref{app:Euler_elastica} we compare the two solutions and find excellent agreement.

\paragraph{Twisted ribbon.}
 {
The nonlinearity of the FvK equations often hinders analytical progress, and most solutions describe developable surfaces with zero Gaussian curvature. 
Contrary to that are twisted ribbons which hold negative Gaussian curvature, and appear in contexts ranging from graphene~\cite{kovsmrlj2016response,hanakata2021thermal}
to self-assembly~\cite{selinger2004shape}.
Here we solve the twisted ribbon problem and show that the CP formulation provides a route for analyzing non-developable configurations.
}

 {
Consider a ribbon of width $W$ and length $L$ whose edges are rotated about the centerline while the end-to-end distance is fixed. 
This problem is remarkably rich, particularly in the  {nearly } isometric limit, where multiple solution branches break translational symmetry along the ribbon~\cite{korte2011triangular,chopin2015roadmap}. 
Here we focus on stretchable ribbons that form helicoidal configurations with stresses and curvatures symmetric along the ribbon.
The solution of Eqs.~\eqref{eq:chi}-\eqref{eq:psi} consistent with these symmetries is}
\begin{eqnarray}
\chi(u,v) = \,\tfrac{Y}{24}\kappa ^2 \, v^{4} \;, \quad \psi(u,v) = \kappa  \, u v,
\label{eq:SolTwist}
\end{eqnarray}
 {with the characteristic curvature $\kappa = {\theta}/{L}$ equal to the imposed twist per unit length.
In the example shown in \figref{fig:all_sims}(c), $h\kappa = 6.28\times10^{-3}$, with numerical accuracy $\varepsilon \sim 10^{-2}$  (see \appref{twisted_derivation} for more general solutions).
}
The reconstructed surface corresponding to~\eqref{eq:SolTwist}, shown in \figref{fig:motivation}(a), exhibits uniform nonzero Gaussian curvature with vanishing mean curvature. 
A comparison with full finite-element simulations is presented in \figref{fig:all_sims}(c): panel (c1) shows the configuration colored by Gaussian curvature, while (c2) displays the difference between the analytic and numerical curvature across the ribbon.

\paragraph{Sliced annulus.}
We now examine an annulus with a radial cut, where the opening angle  {$\tau$, illustrated in Fig.~\ref{fig:motivation}(c)}, is prescribed. 
This geometry is relevant to the mechanics of conical surfaces~\cite{PhysRevLett.101.156104,PhysRevE.91.022404}  {and annular strips~\cite{starostin2022forceless}}, and forms a basic mechanism in kirigami~\cite{moshe2019nonlinear, sadik2021local} and origami~\cite{PhysRevE.94.013002}.
 {We focus here on the isometric limit, in which the SP is set to zero $\chi=0$, and the shape is generated solely by the CP $\psi$ (see \appref{isometry}). The more general case with nonzero SP, is deferred to \appref{open_disc}.}
Assuming curvature only in the azimuthal direction  {to preserve the zero Gaussian curvature of the sheet, the solution  {$ \psi(r,\theta)= c~r~\theta \cos(\theta - \varphi)\, $}  in polar coordinates $(r,\theta)$ provides a valid approximation as long as the azimuthal radius of curvature is large compared with the thickness \footnote{Note that the assumption of curvature only in the azimuthal direction is not the most general one, and is made to simplify the analysis. A more general form would allow the principal curvature to vary in space.}.

The characteristic curvature is $\kappa \sim 1/r$, and the constants $c$ and $\varphi$ are set by stress and torque boundary conditions, with representative configurations shown in \appref{open_disc}. 
Displacement boundary conditions are imposed on the configuration reconstructed numerically from Eq.~\ref{eq: Weingarten}, illustrated in \figref{fig:motivation}(c). 
A comparison between the analytic and finite-element solutions is shown in \figref{fig:all_sims}(b).
Panel (b1) shows the equilibrium configuration colored by mean curvature, while (b2) displays the small difference between analytical and numerical mean curvature across the annulus.
In this example $(h \kappa)$ takes its largest value at the inner boundary with $(h \kappa)=0.05$, and the numerical accuracy is $\varepsilon \sim 10^{-2}$.}

\begin{figure}
    \centering
\includegraphics[width=\columnwidth]{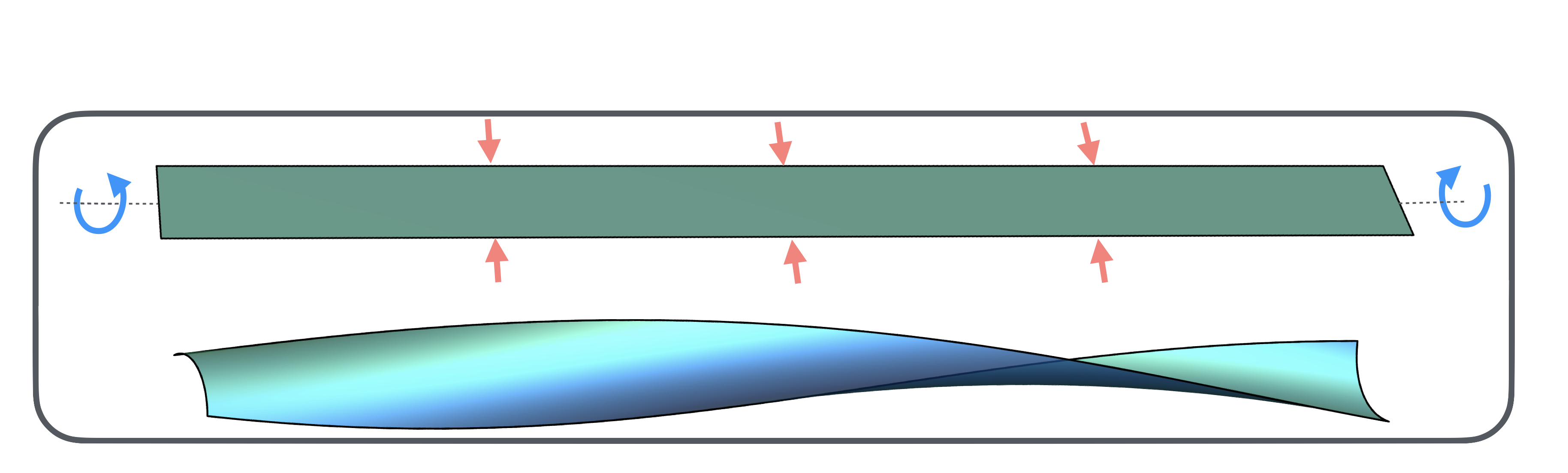}
    \caption{ {\textbf{Mixed loading with hidden intrinsic symmetry.}
Top: Boundary conditions consisting of simultaneous twist and compression, applied across its width. Bottom: Corresponding ribbon configuration reconstructed by integrating the Weingarten equations using the analytical solutions of Eq.~(\ref{eq:MixedLoading}). Although the resulting configurations exhibit no obvious symmetry, the underlying stress and curvature potentials retain simple symmetric forms.}}
    \label{fig:MixedLoading}
\end{figure}

 {\textbf{\textit{Mixed loading.}}
The previous examples demonstrate how the CP formulation extends the FvK theory to multi-valued configurations. Here we illustrate a second advantage of the intrinsic formulation. In the classical height-function formulation, analytical simplifications rely on symmetry assumptions imposed directly on the configuration, thereby constraining the admissible embeddings. In contrast, the CP formulation allows symmetry assumptions to be imposed on the stress and curvature potentials rather than directly on the embedding. Consequently, the reconstructed configuration may remain geometrically complex while the governing equations still reduce to ordinary differential equations. This distinction becomes particularly important when several loading modes are combined.

To illustrate this principle, we consider three representative examples in which compression is superposed with twist and shear. Two cases are presented in the End Matter (Figs.~4 and 5). Here we focus on the case of simultaneous twist and compression in the transverse direction, shown in Fig.~\ref{fig:MixedLoading}. Although the configuration exhibits no obvious symmetry, the boundary conditions suggest simple symmetric ans\"atze in which the stress and curvature depend only on the transverse coordinate, $\chi(u,v)=c_1u^2+c_2uv+\chi_1(v)$ and $
\psi(u,v)=c_3u^2+c_4uv+\psi_1(v)$. 
This reduces Eqs.~(5)-(6) to ordinary differential equations, yielding
\begin{eqnarray}
\chi= - B\lambda^2 u^2/2  + Y \kappa^2 v^4 /24,\quad
 \psi= \kappa \,u \,v  + m \cos(\lambda v)\;.%\notag
\label{eq:MixedLoading}
\end{eqnarray}
The corresponding configuration, obtained by integrating Eq.~\ref{eq: Weingarten}, is shown in Fig.~\ref{fig:MixedLoading}. This demonstrates that the intrinsic nature of the CP formulation extends the analytical power of the FvK theory by shifting the relevant notion of symmetry from the configuration to the intrinsic elastic fields.
}

\textbf{\textit{Summary and Discussion.}}
We introduced an intrinsic curvature-potential formulation of thin-sheet elasticity that bypasses the limitations of weakly nonlinear, graph-based descriptions, {while substantially extending the class of analytically tractable problems}. 
Comparison with full finite-element minimization across several examples shows that the curvature-potential approach remains accurate even under extreme confinement, in regimes where classical FvK-type theories fail.

 {The advantages of the CP formulation are particularly apparent in the statistical mechanics of fluctuating membranes and ribbons, where much of the theoretical progress relies on the nonlinear structure of the FvK equations. First, curvature-based observables retain their classical form; for example,
$\langle \partial_i\mathbf{n}(\mathbf{q})\cdot\partial_j\mathbf{n}(-\mathbf{q})\rangle
=\langle \bar{a}^{kl}b_{ik}(\mathbf{q})\,b_{jl}(-\mathbf{q})\rangle$
holds whether $b$ is represented by a shallow height function or by a curvature potential. Second, removing the graph restriction opens these methods to multi-valued states, including twisted ribbons, that fall outside the conventional height-function description. Third, fluctuation theories generally require a static solution about which fluctuations are expanded. By extending the class of analytically accessible base states, the CP formulation provides a route to studying fluctuations and renormalized elastic properties under mixed loading and other complex boundary conditions. These advances open new directions for the mechanics and statistical physics of systems such as graphene and lipid bilayers, particularly when large geometric deflections coexist with small strains. }

Because the formulation is derived directly from the geometric elastic energy, it extends immediately to sheets with non-Euclidean reference metrics $\bar{a}$ or finite reference curvature $\bar{b}$. 
This includes systems shaped by growth, plastic deformation, residual stress, or programmed geometry, where incompatibility plays a central mechanical role. 
These cases lie beyond the scope of the present work but provide natural directions for future extensions.

 {The curvature-potential } formulation raises new geometric questions concerning the global structure of surfaces. 
In particular, solving the elastic problem on a periodic annular domain requires enforcing not only the local Gauss–MPC compatibility conditions but also global topological constraints that couple the fundamental forms around the periodic direction. 
Deriving these global relations will allow us to address the problem of a dislocated annulus and offer new insight into the finite-energy hypothesis for dislocations~\cite{kupferman2017bending,kupferman2025willmore}.

Overall, the curvature-potential formulation provides a unified geometric framework bridging the gap between weakly nonlinear plate theories and the fully nonlinear geometry of thin sheets. 
It preserves the structure and intuition of classical approaches while extending their reach to multi-valued, large-slope, and non-Euclidean configurations. 
We expect this framework to find broad application in architected materials, growth-driven morphogenesis, kirigami and origami mechanics, and thermalized two-dimensional materials, and to serve as a foundation for further developments in geometric elasticity.

\textbf{Acknowledgments.}
This work was supported by the Israel Science Foundation under Grant No.~1857/26.
M.M. acknowledges the hospitality of the Kavli Institute for Theoretical Physics during the programs Sheets-16 and Active-24, where ideas underlying this work were developed. This research was supported in part by grant NSF PHY-2309135 to the Kavli Institute for Theoretical Physics (KITP). C.M. was supported by ISF 2304/24 and BSF 2022076.

\section*{Data Availability}
The computational notebook used to perform the calculations
reported in this article is available on Zenodo~\cite{zenodo-notebook}.

\bibliography{references}

\clearpage
\setcounter{equation}{0}
\renewcommand{\theequation}{A\arabic{equation}}
\renewcommand{\theHequation}{endmatter.\arabic{equation}}
\section*{End  matter}
\textbf{Additional mixed-loading solutions.—}
The mixed-loading example in the main text demonstrates that the CP formulation can remain analytically tractable even when the reconstructed configuration exhibits no obvious symmetry. Here we present two additional solutions, combining compression with shear and with twist, respectively, together with the accompanying boundary bending moments. Although these combined loads produce geometrically complex configurations, the stress and curvature fields retain a simple structure that reduces the governing equations to analytically solvable ordinary differential equations.

\textit{Compression and shear.}
Consider a ribbon of length $L$ subjected to longitudinal compression $F$ and shear $\Sigma$. The corresponding solution carries an accompanying boundary bending moment $\xi$ and is generated by the potentials
\begin{equation}
\chi=\Sigma uv-\frac{1}{2} B \kappa^2 v^2,
\qquad
\psi=A\cos(\kappa u),
\label{eq:compression-shear}
\end{equation}
where $\kappa = \sqrt{F/B}$, and $A=\xi/\left(F \cos(\kappa\,L/2)\right)$.

At the short edges $u=\pm L/2$, the corresponding stress traction and bending moment are
\begin{equation}
\sigma\cdot\hat{\mathbf n}=\mp(F,\Sigma),
\qquad
m\cdot\hat{\mathbf n}=\mp(\xi,0)\,.
\end{equation}

\begin{figure}[h]
    \centering
    \includegraphics[
        width=\linewidth,
    ]{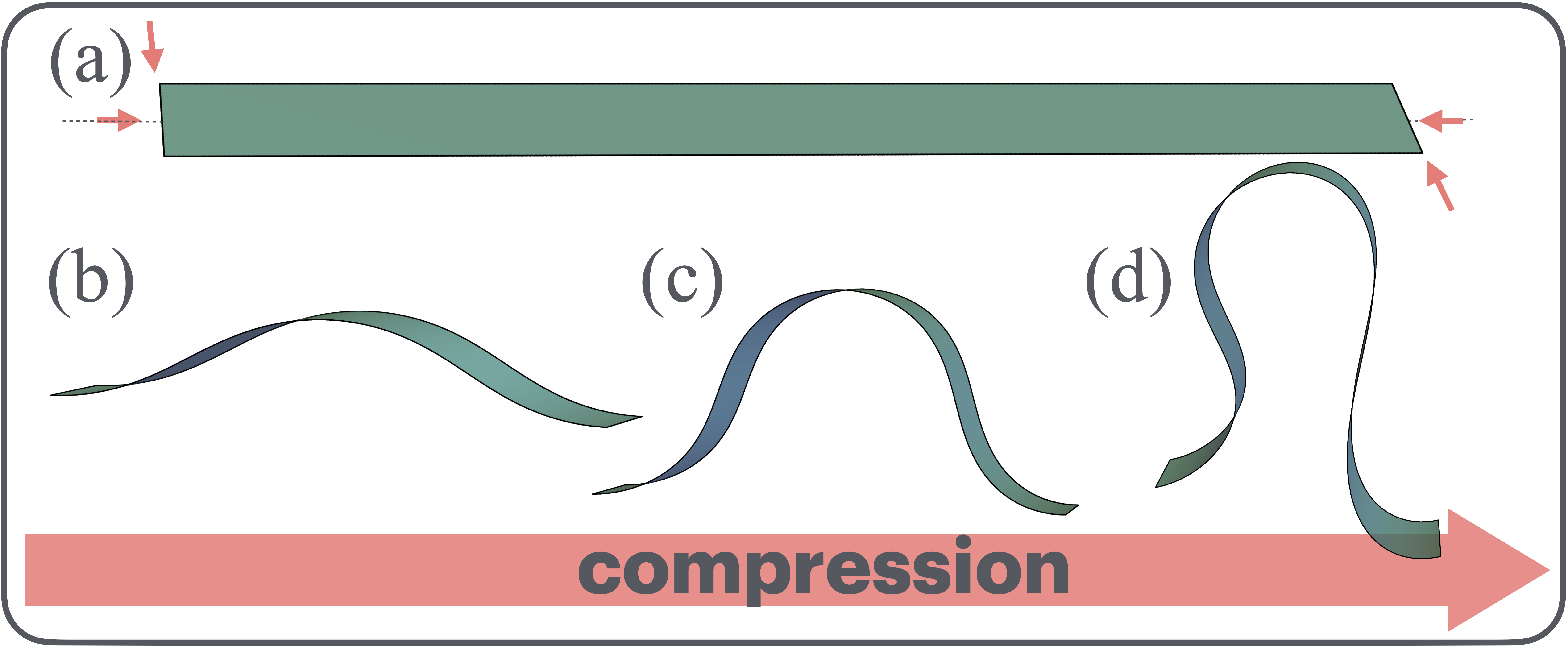}
    \caption{
Ribbon subjected to shear and compression.
(a) Loading protocol: shear force and compression are applied on the short edges.
(b-d) Reconstructed configurations for fixed shear and increasing compression.
}
\label{fig:end_shear_comp}
\end{figure}

The resulting configurations for fixed shear ($\Sigma$) and increasing compressive force ($F$), and consequently moment ($\xi$), are illustrated in \figref{fig:end_shear_comp}.

\begin{figure}[h]
    \centering
    \includegraphics[
        width=\linewidth]{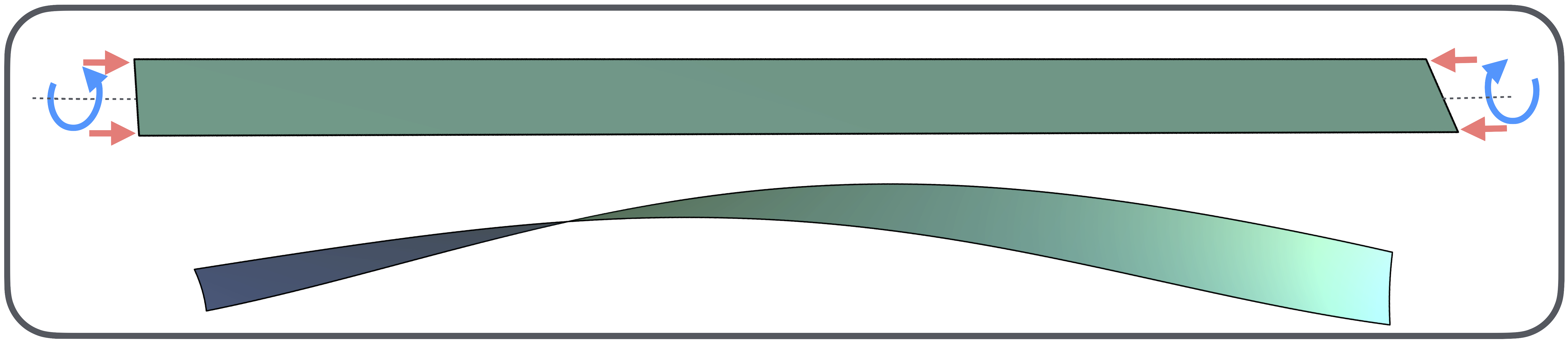}
    \caption{
Ribbon subjected to compression, twist and bend.
Top: loading protocol, with compression and twisting moments applied on the short edges.
Bottom: reconstructed configuration obtained from the CP solution.
}
    \label{fig:end_twist_comp}
\end{figure}

\textit{Compression, twist and bend.}
Consider a ribbon subjected simultaneously to longitudinal compression $F$, twist $\tau$ and bend $\eta$. An analytical solution is
\begin{equation}
\chi=-\frac{F}{2}v^2+\frac{Y\tau^2}{24}u^4,
\qquad
\psi=\tau uv+A\cos(\kappa u),
\label{eq:compression-twist}
\end{equation}
where
\begin{equation}
\kappa=\sqrt{\frac{F}{B}},
\qquad
A=\frac{\eta}{F\cos(\kappa L/2)}.
\end{equation}
The corresponding boundary traction and moment are
\begin{eqnarray}
\left.\sigma\cdot\hat{\mathbf n}\right|_{u=\pm L/2}
&=&\mp(F,0)\,, \\
\left.m\cdot\hat{\mathbf n}\right|_{u=\pm L/2}
&=&
\left(\mp\eta,\,
\pm B(1-\nu)\tau\right).
\label{eq:compression-twist-bc}
\end{eqnarray}
The reconstructed configuration is shown in \figref{fig:end_twist_comp}.

Together with the example in the main text, these solutions demonstrate that mixed loading can preserve a simple analytical structure in the stress and curvature fields even when the reconstructed configuration exhibits no obvious symmetry.

\newpage

\newcommand{\est}{\epsilon_{\mathrm{st}}}
\newcommand{\ebn}{\epsilon_{\mathrm{bn}}}
\renewcommand{\paragraph}[1]{\textbf{\textit{#1}}}
\newcommand{\Revised}[1]{\noindent \color{red}  #1\normalcolor}
\onecolumngrid

{\center{\Large\textbf{{Supplementary Material}}}}

\section{Numerical methods}\label{numerical_methods}

To determine energy-minimizing configurations of elastic sheets with prescribed
reference metric $\bar a$ and reference curvature $\bar b$, we perform a finite
element simulation based on a triangulated discretization of the surface. The sheet
is represented as a two-dimensional mesh of linear triangular elements embedded in
$\mathbb{R}^3$. Each triangle carries prescribed reference tensors $\bar a$ and
$\bar b$, which encode the locally preferred in-plane geometry and curvature.

For each triangular face we reconstruct the intrinsic metric $a$ from the
current edge lengths. Let $\mathbf{X}_1,\mathbf{X}_2,\mathbf{X}_3\in\mathbb{R}^2$
denote the reference coordinates of the triangle vertices and define the
reference edge vectors $\mathbf{s}_{ij}=\mathbf{X}_j-\mathbf{X}_i$. Writing the
metric tensor as
$a=\begin{pmatrix}a_{11} & a_{12}\\ a_{12} & a_{22}\end{pmatrix}$,
the squared length of a tangent vector $\mathbf{s}=(s_x,s_y)$ in this metric is
\begin{eqnarray}
\|\mathbf{s}\|_a^2 = a_{11}s_x^2 + 2a_{12}s_x s_y + a_{22}s_y^2 .
\end{eqnarray}
Let $l_{ij}$ denote the current Euclidean length of the edge between vertices
$i$ and $j$ in $\mathbb{R}^3$. Requiring that the metric reproduces the squared
lengths of the three triangle edges,
$\|\mathbf{s}_{ij}\|_a^2 = l_{ij}^2$, yields the linear system
\[
\begin{pmatrix}
(s_{23}^x)^2 & 2 s_{23}^x s_{23}^y & (s_{23}^y)^2 \\
(s_{31}^x)^2 & 2 s_{31}^x s_{31}^y & (s_{31}^y)^2 \\
(s_{12}^x)^2 & 2 s_{12}^x s_{12}^y & (s_{12}^y)^2
\end{pmatrix}
\begin{pmatrix}
a_{11}\\ a_{12}\\ a_{22}
\end{pmatrix}
=
\begin{pmatrix}
l_{23}^2\\ l_{31}^2\\ l_{12}^2
\end{pmatrix},
\]
whose solution determines the discrete metric tensor $a$ for the triangle.

\begin{figure}[h]
    \centering
    \includegraphics[width=0.4\linewidth,trim=55 70 60 80,clip]{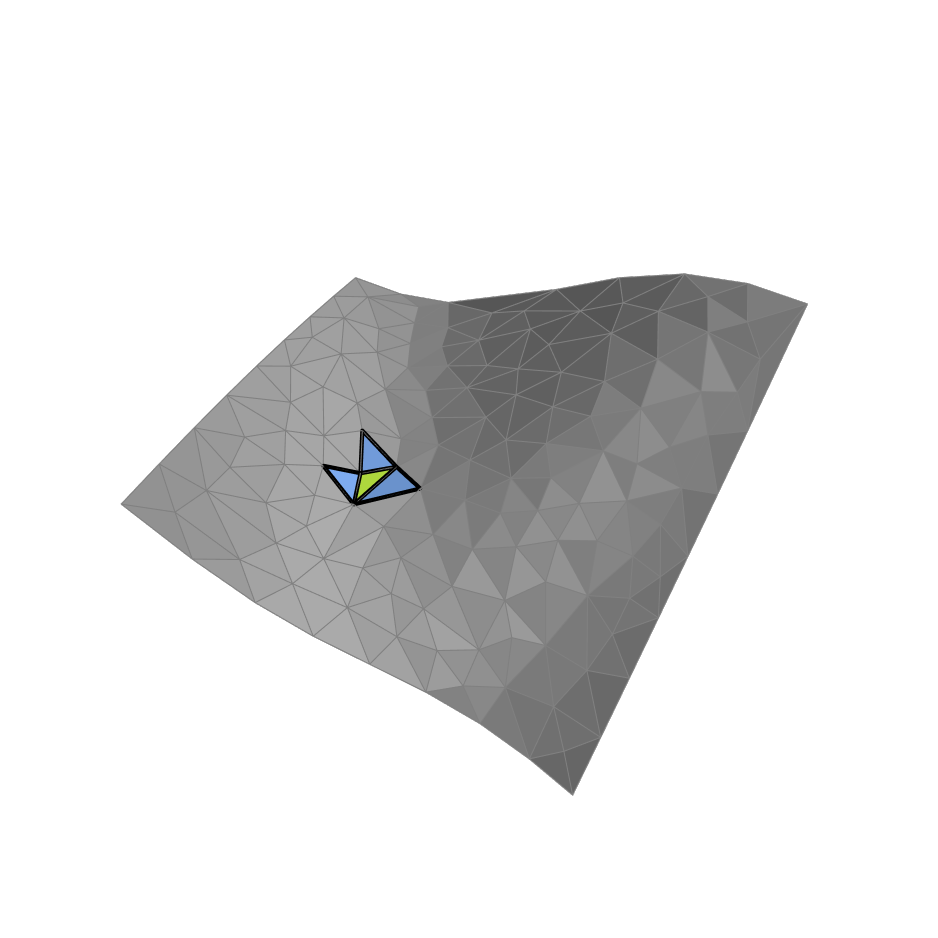}
    \hspace{2pt}
    \includegraphics[width=0.4\linewidth,trim=300 150 200 200,clip]{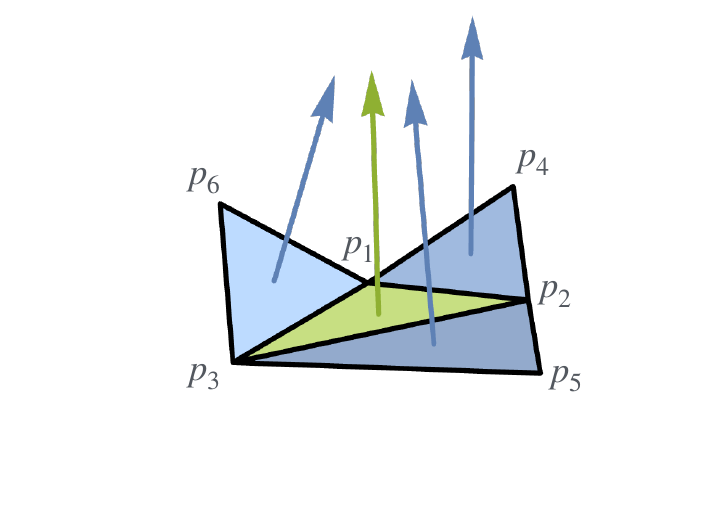}
   \caption{
Local geometric reconstruction on the triangular mesh.
Left: triangulated surface with a triangle and its neighboring triangles highlighted.
Right: stencil used to compute the curvature tensor $b$. The central triangle and its neighbors define six vertices $p_i$. The arrows indicate the surface normals; the normal of the central triangle defines the reference plane used to measure vertex heights and reconstruct the curvature tensor $b$.
}
    \label{fig:numeric}
\end{figure}

The curvature tensor $b$ is reconstructed from the current vertex positions
using the three vertices of the triangle together with their three
edge-adjacent neighboring vertices. Let
$\mathbf{p}_i\in\mathbb{R}^3$, $i=1,\dots,6$, denote these vertices
(see \figref{fig:numeric}), and let $\mathbf{X}_i\in\mathbb{R}^2$
denote their reference coordinates.

The reference face center is
$\mathbf{X}_0=\tfrac13(\mathbf{X}_1+\mathbf{X}_2+\mathbf{X}_3)$
And the 3D face center and unit normal are
\[
\mathbf{P}_0=\tfrac13(\mathbf{p}_1+\mathbf{p}_2+\mathbf{p}_3), \qquad
\mathbf{n}_0=\frac{(\mathbf{p}_2-\mathbf{p}_1)\times(\mathbf{p}_3-\mathbf{p}_1)}
{\|(\mathbf{p}_2-\mathbf{p}_1)\times(\mathbf{p}_3-\mathbf{p}_1)\|}.
\]

Each vertex provides a signed height relative to the face plane,
$h_i=(\mathbf{p}_i-\mathbf{P}_0)\cdot\mathbf{n}_0$. 
We fit the quadratic height function:
\begin{eqnarray}
z(x,y)=c_0+c_1 x+c_2 y+\tfrac12 Lx^2+Mxy+\tfrac12 Ny^2
\end{eqnarray}
to the six vertices by imposing for each one of these vertices
$h_i=z(x_i,y_i)$.
Solving the resulting $6\times6$ linear system yields the curvature tensor 
$b=\begin{pmatrix}L & M \\ M & N\end{pmatrix}$.

With the metric and curvature tensors available on each triangle, the local
stretching and bending energy densities are computed by comparing $a$ and $b$ with
the prescribed reference tensors $\bar a$ and $\bar b$ as in Eq.~(1) of the main text. The total elastic energy is
the sum of all triangle contributions weighted by their reference areas.

The resulting discrete energy is a smooth function of the vertex positions.
Its gradient is assembled by differentiating the metric and curvature
reconstruction with respect to the vertex coordinates. Energy minimization is
performed using the Fletcher--Reeves nonlinear conjugate-gradient method
(GSL implementation) with analytic gradients. The optimizer is initialized
with step size $10^{-2}$ and employs a line search with tolerance $10^{-6}$;
convergence is declared when the Euclidean norm of the energy gradient falls
below $10^{-6}$.

The simulation checks each triangle to ensure that it does not collapse or become excessively flat. If a triangle approaches an invalid configuration, the program prints a warning. Triangle inversion is not treated separately; configurations approaching inversion lead to large elastic energies and typically prevent convergence of the minimization. In practice, the line-search globalization of the conjugate-gradient method ensures a monotonic decrease in the energy and prevents severe element distortion in the simulations reported here.

\subsection*{mesh dependency}
\begin{figure}[h]
    \centering
    \includegraphics[width=0.7\linewidth,
    trim=45 15 10 15, clip]{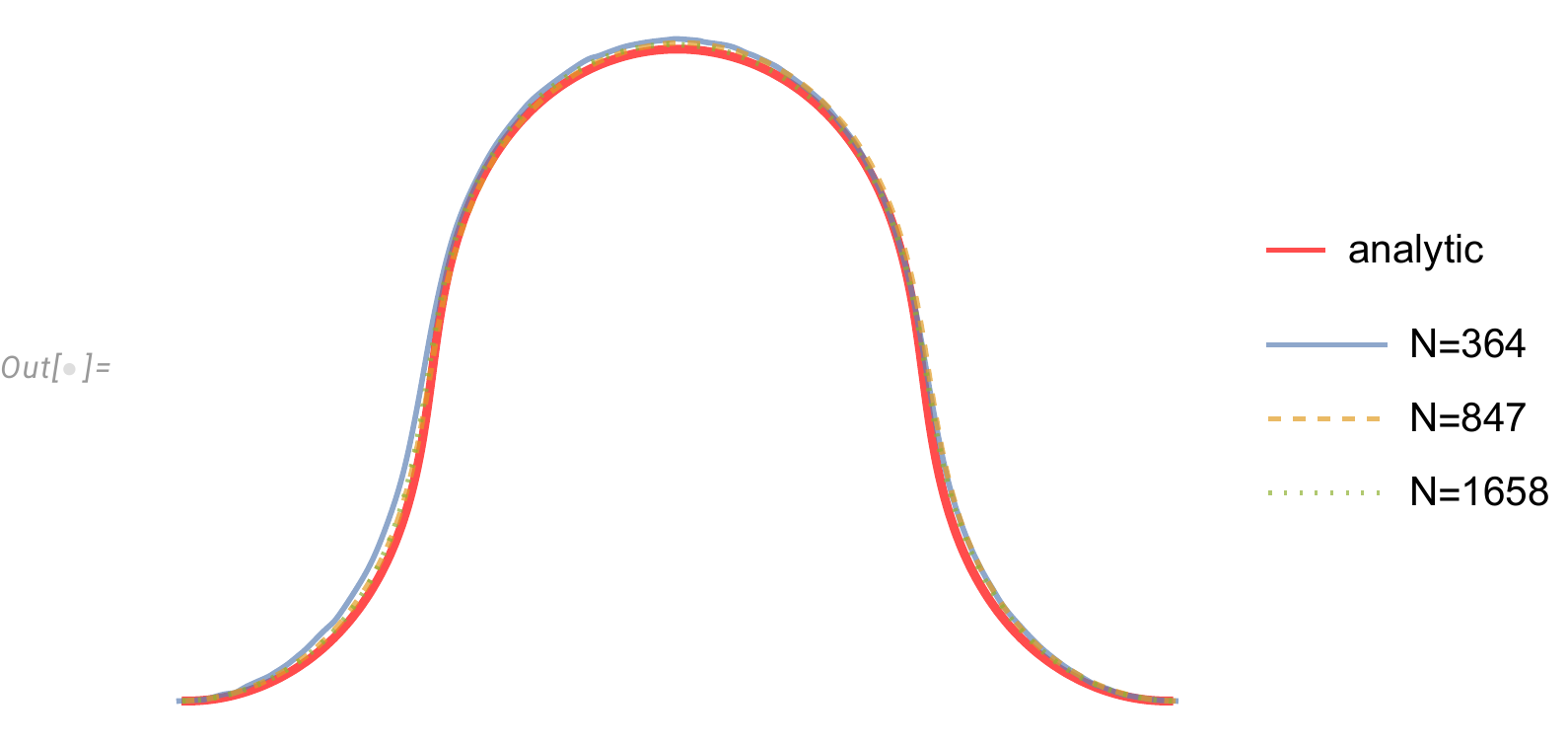}
    \caption{A ribbon of dimensions $10\times 1$ thickness $h=0.1$ Young's modulus $Y=100$ and Poisson's ratio $\nu=0.3$ with compression of $\frac{\Delta L}{L}=0.54936$ (compression defined as in Fig.~1 of the main text). Each color represents the final midline configuration with $N$ the number of vertices in the mesh of the finite element simulation. The red line is the analytic CP solution. }
    \label{fig:mesh_size}
\end{figure}

To assess the dependence of the simulated configuration on mesh resolution, we took a compressed ribbon simulation and repeated the simulations with different numbers of vertices $N$ (here $N\in\{364,847,1658\}$). 
Then we extracted a centerline as illustrated in \figref{fig:mesh_size}.

To quantify the difference between two curves, we compute the root mean square (RMS) Euclidean distance between corresponding sampled points. 
Each curve is first reparameterized by arc length and uniformly sampled at $N_s=1000$ points. 
Given two sampled curves 
$\{p_i\}_{i=1}^{N_s}$ and $\{q_i\}_{i=1}^{N_s}$ in $\mathbb{R}^2$, the RMS distance is defined as
\[
d_{\mathrm{RMS}}(p,q) = 
\sqrt{\frac{1}{N_s}\sum_{i=1}^{N_s} \|p_i - q_i\|_2^2},
\]
where $\|\cdot\|_2$ denotes the Euclidean norm. 
This metric measures the average geometric deviation between the curves and
converges to nearly the same values as the number of sampling points increases.

Table~\ref{tab:rmsdist} reports the pairwise RMS distances between the analytic solution and the curves extracted from three simulations with different mesh resolutions. 
Each simulation is labeled by the number of vertices $N$ in the corresponding mesh.

\begin{table}[h]
\centering
\begin{tabular}{ccc}
\toprule
Comparison & Mesh sizes $(N)$ & RMS distance \\
\midrule
\\
Analytic vs simulation & $364$  & 0.0338785 \\
Analytic vs simulation & $847$  & 0.0235643 \\
Analytic vs simulation & $1658$ & 0.0303141 \\
\midrule
simulation vs simulation & $847,\;364$  & 0.0435472 \\
simulation vs simulation & $847,\;1658$ & 0.0254391 \\
simulation vs simulation & $364,\;1658$ & 0.0404259 \\
\bottomrule
\end{tabular}
\caption{Pairwise RMS distances between the analytic centerline and the centerlines extracted from simulations at three mesh resolutions, and between the simulated centerlines themselves. Each curve is reparameterized by arc length and uniformly sampled at $N_s=1000$ points prior to comparison.}
\label{tab:rmsdist}
\end{table}

\subsection*{Numerical accuracy and error measures}

 {
To quantify the agreement between the analytical predictions and the finite-element simulations, we evaluate the relative energy error
\begin{equation}
\epsilon=\frac{|E_{\mathrm{sim}}-E_{\mathrm{an}}|}{E_{\mathrm{an}}}.
\end{equation}
For the twisted ribbon (Fig.~2(c1) of the main text) we obtain $\epsilon=1.17\times10^{-2}$, while for the clamped ribbon (Fig.~2(a1) of the main text)  at $\Delta L/L=0.7$ we obtain $\epsilon=3\times10^{-2}$. 
For the open annulus (Fig.~2(b1) of the main text), the analytic solutions we presented were for the isometric limit. Therefore, we compare only the bending energy and find $\epsilon=8\times10^{-3}$. 
To quantify the deviation from the reference metric we instead compute the relative $L^2$ error of the metric tensor,
\begin{equation}
\epsilon_a=
\sqrt{
\frac{\sum_i \|a_i^{\mathrm{sim}}-a_i^{\mathrm{an}}\|_F^2}
{\sum_i \|a_i^{\mathrm{an}}\|_F^2}
},
\end{equation}
where the sum runs over all triangles and $\|\cdot\|_F$ denotes the Frobenius norm,
$\|a\|_F^2=\sum_{\alpha,\beta}a_{\alpha\beta}^2$.
For the open annulus this yields $\epsilon_a=2.36\times 10^{-2}$.
}

\section{Deriving the equilibrium equations}\label{full_eq}
To present the complete form of the equilibrium equations as in~\cite{EFRATI2009762}, 
we assume a reference metric $\bar{a}$, and reference curvature tensor $\bar{b}$. The stretching and bending strain tensors are then defined by $ \est = \frac{1}{2}\,(a - \bar{a})$ and $\ebn=\tfrac{1}{2\sqrt{3}}(b - \Bar{b})$.
Based on this, we introduce the stretching and bending stresses, as well as defining the 
total stress $\Sigma$:
\begin{eqnarray*}
    \sigma^{\alpha \beta} &=&  \frac{\partial E}{\partial {(\est)}_{\alpha\beta}} = h A^{\alpha \beta \gamma \delta} {(\est)}_{\gamma\delta} \\
    m^{\alpha \beta} &=& \frac{\partial E}{\partial {b}_{\alpha\beta}} = \frac{h^3}{2\sqrt{3}} A^{\alpha \beta \gamma \delta} (\ebn)_{\gamma\delta}\\
\Sigma^{\alpha\beta} &=& \sigma^{\alpha \beta} + m^{\mu \beta} a^{\gamma \alpha} b_{\mu \gamma}
\end{eqnarray*}

$\nabar$ is the covariant derivative in terms of $\bar{a}$ (marked above as $\text{div}_{\abar}$),
further defining the operator which takes into account both the metric and reference metric meaning $\div$ as \cite{EFRATI2009762} :
\begin{eqnarray}
\left(\div\right)_\mu T^{\mu\nu}&=\nabar_\mu T^{\mu\nu}+\left(\Gamma^\nu_{\alpha\beta}-\Bar{\Gamma}^\nu_{\alpha\beta}\right)T^{\alpha\beta}
\end{eqnarray}
The full energy functional is of the form Eq.~(1) of the main text.
Minimizing in terms of the configuration results with the equilibrium equations Eq.~(3) of the main text (see also \cite{EFRATI2009762}):
\begin{eqnarray}
0&=&(\div)_\beta\left(\Sigma^{\alpha \beta}\right)+(\div)_\beta\left(m^{\mu \beta}\right)a^{\gamma \alpha} b_{\gamma \mu}
\label{efis_eq1SM}\\
0&=&\nabar_\alpha\left((\div)_\beta(m^{\alpha\beta})\right)-\Sigma^{\alpha\beta}b_{\alpha\beta}\label{efis_eq2SM}
\end{eqnarray}

Next, we will look at the leading orders of these equations in the case where $\bar{b} = 0$.
We note the Christoffel symbols term scales like the strain: $\Gamma-\Bar{\Gamma}\sim \est$.
In the FvK scaling regime, where the stretching and bending energies scale
similarly, $\est\sim h^2\kappa^2$ and $\ebn\sim h\kappa^2$, with the
characteristic curvature $\kappa$ determined by the specific system, the
equilibrium equations reduce to:
\begin{eqnarray}
0&=&\nabar_\beta
(\sigma^{\alpha\beta})\label{eq:fvk1}\\
0&=&\nabar_\alpha\nabar_\beta m^{\alpha \beta}-
\sigma^{\alpha \gamma}b_{\alpha \gamma}\label{eq:fvk2}
\end{eqnarray}

Allowing for a general reference curvature $\bar{b}\neq0$ requires returning to the
full equilibrium equations, Eqs.~\eqref{efis_eq1SM} and \eqref{efis_eq2SM}. When
simplifying these equations under various scaling assumptions, it is important to
note that large bending can occur even when the bending strain $\ebn$ remains small.
Consequently, the equilibrium equations acquire additional contributions involving
$\bar{b}$.

Previous work addressed such situations by prescribing a reference height profile
for the sheet \cite{kovsmrlj2013mechanical}. In contrast, a general reference
curvature $\bar{b}$ does not necessarily correspond to a compatible surface
embedding. The treatment of such compatibility issues lies beyond the scope of the
present work and is left for future study.

\section{Boundary conditions and configuration reconstruction}\label{b_t}

\subsection*{stress and torque boundary conditions}

The transition from the classical FvK height function $f$ to the curvature
potential $\psi$ in the geometric formulation introduces similar difficulties
to those encountered when replacing the in-plane displacement field with the
Airy stress potential $\chi$: in both cases the physical boundary conditions
are no longer expressed directly in terms of the fundamental variables.

The free-boundary terms of the equilibrium equations
Eqs.~\eqref{efis_eq1SM}–\eqref{efis_eq2SM} were derived in
\cite{EFRATI2009762} and read
\begin{eqnarray}
0&=&\Sigma^{\mu\nu}n_\nu \label{s_n}\\
0&=&m^{\alpha\beta}n_\alpha n_\beta \label{mnn}\\
0&=&\left(\text{div}_{a,\bar{a}}\right)_\alpha(m^{\alpha\beta})n_\beta .
\label{dmn}
\end{eqnarray}
$n$ is the outward unit boundary conormal, tangent to the surface.
These conditions correspond to vanishing boundary force
(\ref{s_n}) and curvature torques (\ref{mnn})–(\ref{dmn}).

Under the FvK scaling ($\est \sim h\,\ebn$), retaining the leading-order
terms in each equation gives
\begin{eqnarray}
0&=&\sigma^{\mu\nu}n_\nu \label{lins_n}\\
0&=&m^{\alpha\beta}n_\alpha n_\beta \label{linmnn}\\
0&=&(\nabar_\alpha m^{\alpha\beta})n_\beta .
\label{lindmn}
\end{eqnarray}

\subsection*{configuration reconstruction and direct boundary conditions}
An alternative approach is to impose boundary conditions directly on the
configuration of the surface rather than on the boundary forces and torques.
The configuration $\boldsymbol{\phi}(u,v)$ can be reconstructed from the
first and second fundamental forms using the Weingarten equations
\begin{eqnarray}
\partial_\mu \partial_\nu\boldsymbol{\phi}
&=&
\Gamma^\alpha_{\mu\nu}\partial_\alpha\boldsymbol{\phi}
+
b_{\mu\nu}\hat{n}, \\
\partial_\mu\hat{n}
&=&
-a^{\alpha\beta}b_{\mu\alpha}\partial_\beta\boldsymbol{\phi}.
\end{eqnarray}

These equations form a system of linear PDEs for the tangent vectors
$\partial_\mu\boldsymbol{\phi}$ and the normal vector $\hat{n}$.
We solve them either analytically or numerically using the
\texttt{NDSolve} function in \textit{Mathematica}. For arbitrary $a$
and $b$ the solution is generally path-dependent. However, when $a$
and $b$ satisfy the Gauss and Mainardi–Codazzi–Peterson compatibility
conditions, the solution becomes path-independent and the configuration
is uniquely determined up to rigid motions.

When solving the equations numerically we monitor standard geometric
consistency checks. In particular, we verify that the reconstructed normal remains unit and orthogonal to the tangents within numerical tolerance, and that the metric and curvature fields computed from the tangent vectors and from the reconstructed surface
match the prescribed fields obtained from the stress and curvature
potentials.

\paragraph{Compressed ribbon boundary conditions.}\label{clamped_b_c}

As an example, we derive boundary conditions analytically for the
compressed ribbon configuration. In this case the solution admits a
uniform strain, so the metric $a$ is constant and the Christoffel
symbols vanish ($\Gamma^\alpha_{\mu\nu}=0$).
Since the solution has no dependence on the transverse coordinate
$v$ along the short side of the ribbon, we take
$\psi(u,v)=\psi(u)=A\cos(\kappa u)$.
The Weingarten equations
therefore reduce to
\begin{eqnarray}
\partial_u\hat{n} &=& -\psi''\,\partial_u\boldsymbol{\phi},
\label{Weingarten_c1}\\
\partial_u^2\boldsymbol{\phi} &=& \psi''\,\hat{n}.
\end{eqnarray}
This leads to the ODE
\begin{eqnarray}
\hat{n}''(u)
-
\frac{\psi'''(u)}{\psi''(u)}\hat{n}'(u)
+
(\psi''(u))^2\hat{n}(u)=0 .
\end{eqnarray}

For the configuration in $\mathbb{R}^3$ 
we 
impose clamped boundary conditions 
$\hat{n}(\pm L/2)=(0,0,1)$ and the lowest mode $\kappa=\frac{2\pi}{L}$.
Solving the equation gives
\[
\hat{n}(u)=
\left(
\sin\!\left(\kappa {A} \sin(\kappa u)\right),
0,
\cos\!\left(\kappa {A}\sin(\kappa u)\right)
\right).
\]

Using Eq.~\eqref{Weingarten_c1}, the distance between the two ribbon
edges $L-\Delta L=\big|\boldsymbol{\phi}(L/2,v)-\boldsymbol{\phi}(-L/2,v)\big|$ is given by:
\begin{eqnarray}
\boldsymbol{\phi}(L/2,v)-\boldsymbol{\phi}(-L/2,v)
=
\int_{-L/2}^{L/2}\partial_u\boldsymbol{\phi}\,du
=
\int_{-L/2}^{L/2}-\frac{\partial_u\hat{n}}{\psi''}\,du
\end{eqnarray}
And results with 
\begin{equation}
   1-\frac{\Delta L}{L}=
J_0\!\left(\frac{2\pi A}{L}\right)
\end{equation}
where $J_0$ is the Bessel function of the first kind.
This expression gives the distance between the short edges of the
ribbon and allows us to impose the boundary conditions through the
parameter $A$.

\section{The isometric limit}\label{isometry}
There exist configurations in which the deformation is purely bending,
so that the metric satisfies $a=\bar{a}$. Such isometric solutions
are particularly relevant in the thin limit $h\to0$, where stretching
is energetically much more costly than bending.

In this regime the problem reduces to minimizing the bending energy (Eq.~(1) of the main text) subject
to the compatibility conditions (Eq.~(2) of the main text). Writing the curvature tensor
$b$ in terms of the curvature potential (CP) automatically satisfies one of
these conditions, while the remaining one is imposed using a Lagrange
multiplier. The resulting functional is
\begin{eqnarray}
U=\intop\left[\frac{h^{3}}{24}A^{\alpha\beta\gamma\delta}b_{\alpha\beta}b_{\gamma\delta}
-\lambda\left(\det\left(\frac{b}{\bar{a}}\right)-\bar{K_{G}}\right)\right]dS,
\qquad
b_{\alpha\beta}=\nabar_{\alpha\beta}\psi .
\end{eqnarray}

Varying the functional with respect to the curvature potential $\psi$
yields the bulk equation:
\begin{eqnarray}
B\bar{\Delta}\bar{\Delta}\psi=\Lambda \cdot b ,
\qquad
\Lambda^{\alpha\beta}=
\frac{\epsilon^{\alpha\mu}}{\sqrt{\bar{a}}}
\frac{\epsilon^{\beta\nu}}{\sqrt{\bar{a}}}
\nabar_{\mu\nu}\lambda .
\end{eqnarray}

Here $\lambda$ plays a role analogous to the stress potential $\chi$.
In addition, variation with respect to $\lambda$ enforces the geometric
constraint
$\det(\bar{a}^{-1}b)=\bar{K}_G $.

\section{Compressed ribbon}\label{compressed_ribbon}

For a ribbon with
dimension 
$\left(u,v,w\right)\in\left[-\frac{L}{2},\frac{L}{2}\right]\times\left[-\frac{W}{2},\frac{W}{2}\right]\times\left[-\frac{h}{2},\frac{h}{2}\right]$
where $h\ll W\ll L$ with flat reference terms $\bar{a}=\text{Id},\bar{b}=0$,
the general solutions which solve the equilibrium equations in the FvK limit are \cite{AudolyPomeau2010-Ch8}:
\begin{eqnarray}\label{eq:shear_chi_psi}
\chi(u,v)= - \frac{Y h^2}{24(1-\nu^2)} \kappa^2 v^{2}+c_1 u v+c_2 u^2,\qquad
\psi(u,v)=A\cos(\kappa u)+B\sin(\kappa u),
\end{eqnarray}
These solutions yield constant stress and have the same shape in the isometric limit where the SP $\chi$ takes the position of a Lagrange multiplier insuring zero Gaussian curvature (see \appref{isometry}). 

$A,B,\kappa$ must obey the boundary condition, in our case we force the short edges of the ribbon to stay flat meaning $B=0$ and $\kappa=\frac{2\pi}{L}n$. For small loads we expect to find the first mode-$n=1$.
$A$ is then determined by the distance between the short edges (see \appref{clamped_b_c}).
\begin{figure}[h]
    \centering
    \includegraphics[width=0.8\linewidth,trim=5 230 5 210,clip]{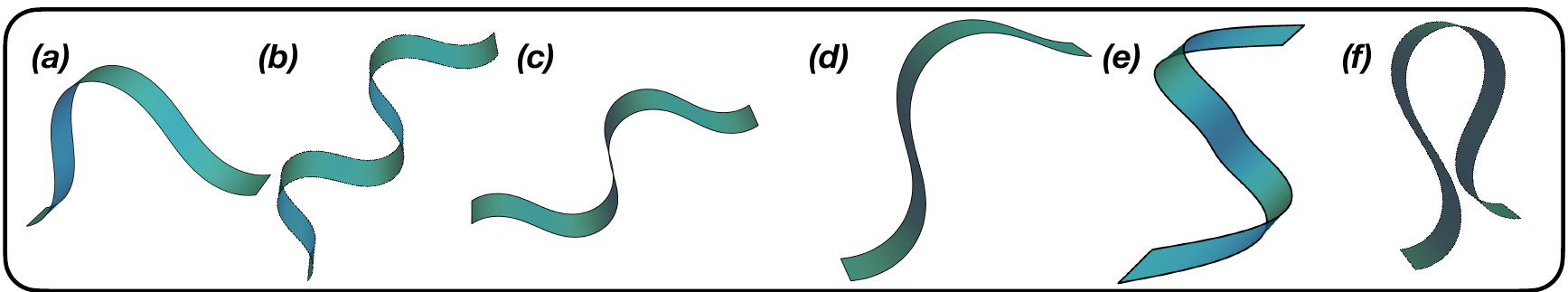}
    \caption{
The configurations correspond to a ribbon of dimensions $10\times 1$, reconstructed from the Weingarten equations with $\chi$ and $\psi$ as in~\eqref{eq:shear_chi_psi}, with $c_2 = 0$. 
(a–d) Configurations without shear ($c_1 = 0$): 
(a) $\kappa = \tfrac{2\pi}{10},~ A = 2,~ B = 0$; 
(b) $\kappa = \tfrac{4\pi}{10},~ A = 1,~ B = 0$; 
(c) $\kappa = \tfrac{3\pi}{10},~ A = 0,~ B = 1.25$; 
(d) $\kappa = \tfrac{2\pi}{10},~ A = 1.25,~ B = 1.25$; 
(e,f) Configurations with shear: 
(e) $\kappa = \tfrac{2\pi}{10},~ A = 2,~ B = 3,~ c_1 = 3$; 
(f) $\kappa = \tfrac{2\pi}{10},~ A = 3,~ B = 1,~ c_1 = 1$.
}

    \label{fig:shear}
\end{figure}

 {Allowing shear ($c_1$) as well and higher modes ($\kappa$) results with a large variety of configurations. These configurations aren't simple height functions, some examples are represented in \figref{fig:shear} constructed via  Weingarten equations. These configurations follow many of the experimental configurations found in \cite{yu2019bifurcations}.}

\subsection*{Euler elastica comparison}\label{app:Euler_elastica}

The planar Euler elastica is a one-dimensional variational problem for the
centerline of an inextensible rod parameterized by arc length $s\in[0,L]$.
The configuration of the rod is described by a planar curve $(x(s),y(s))$.
Let $\theta(s)$ denote the tangent angle of the curve. The curvature is
$\kappa=\theta'(s)$, and the unit tangent vector is
$(x'(s),y'(s))=(\cos\theta,\sin\theta)$.

For bending stiffness $B$, the bending energy is
$E=\frac{B}{2}\int_0^L (\theta')^2\,ds$.
Imposing a prescribed end-shortening $\delta$ corresponds to the geometric
constraint
$\int_0^L \cos\theta(s)\,ds = L-\delta$.

This constraint is incorporated variationally via a Lagrange multiplier
$\lambda$, leading to the  functional
\begin{equation}
\mathcal{F}[\theta]
=
\int_0^L
\left[
\frac{B}{2}(\theta')^2
+\lambda\big(\cos\theta - (1-\delta/L)\big)
\right]ds 
\end{equation}
where $\lambda$ enforces the horizontal end-to-end distance. This yields the Euler-Lagrange equation
\begin{equation}
B\,\theta''(s)+\lambda\sin\theta(s)=0,
\end{equation}

\begin{figure}[h]
    \centering
    \includegraphics[width=0.95\linewidth]{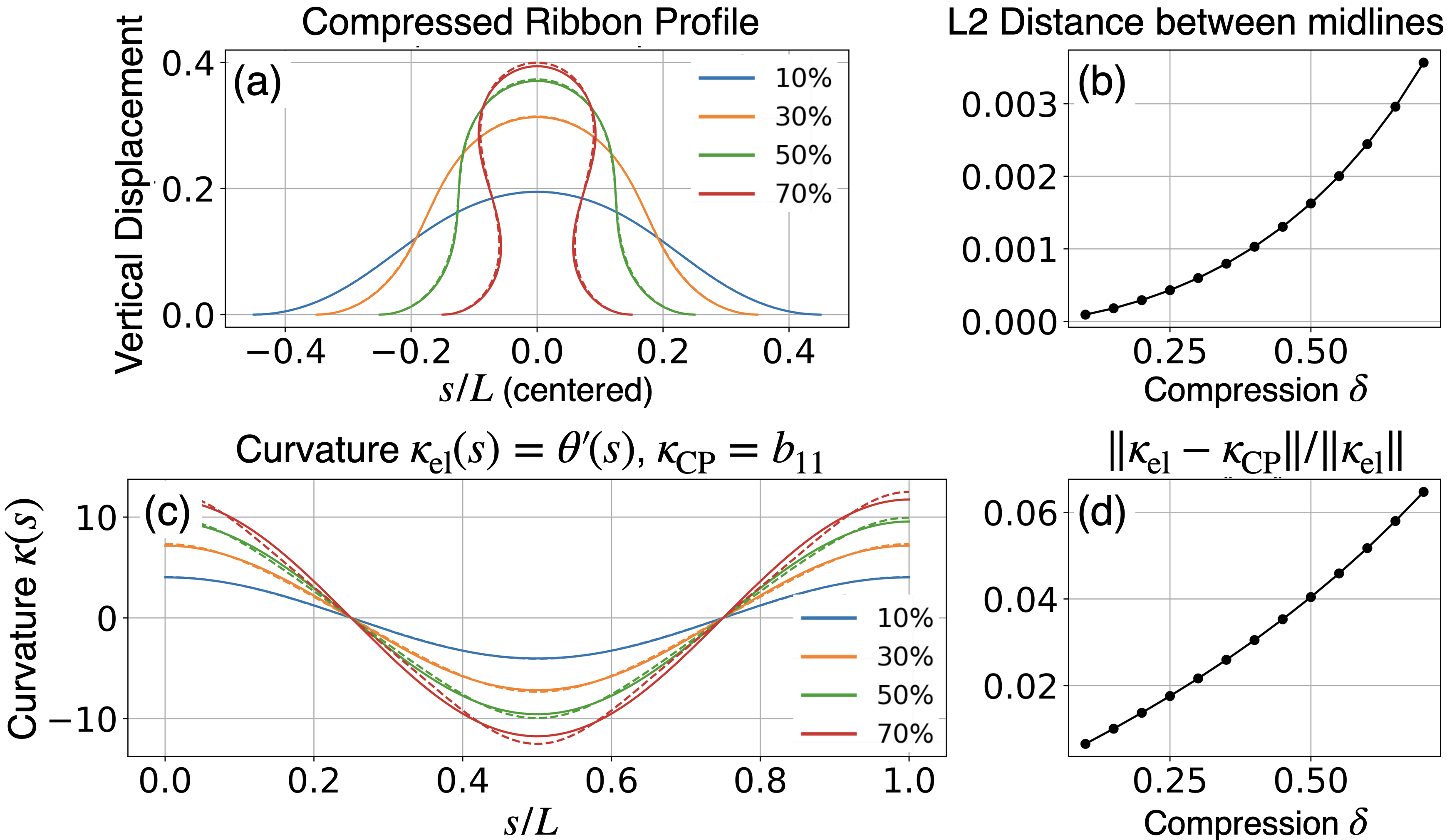}
    \caption{Comparison between classical Euler elastica configurations (dashed) and curvature-potential (CP/Weingarten) solutions (solid) for increasing compression $\delta \in \{10\%,30\%,50\%,70\%\}$. 
Top left: midline profiles configuration of the curvature potential solutions compared with the Euler elastica solutions. 
Top right: $L^2$ distance between the two midlines as a function of compression. 
Bottom left: curvature comparison, showing $\kappa_{\mathrm{el}}(s)=\theta'(s)$ and the CP curvature along the ribbon (from the curvature tensor $b_{ij}$) $\kappa_{\mathrm{CP}}=b_{11}$. 
Bottom right: relative $L^2$ curvature differences $\|\kappa_{\mathrm{el}}-\kappa_{\mathrm{CP}}\|/\|\kappa_{\mathrm{el}}\|$ as a function of compression. 
}
    \label{fig:euler_elastica}
\end{figure}

Figure~\ref{fig:euler_elastica}(a) compares midlines and curvatures obtained from the classical Euler elastica with those generated by the curvature-potential (CP) formulation under identical end-shortening $\delta$. 
At small compressions, the two models produce nearly indistinguishable midlines and curvature distributions. 
As compression increases, systematic differences emerge, although they remain quantitatively small, the distance difference is shown in \figref{fig:euler_elastica}(b). 
For an intrinsic comparison, we directly evaluate the curvature fields and quantify their discrepancy via the $L^2$ difference and the relative error shown in the Fig.~\ref{fig:euler_elastica}(c-d).

 {It is important to note that in the strictly isometric limit, the curvature-potential formulation is equivalent to Euler elastica. Upon solving the highly simplified Weingarten equations, the curvature potential in this case satisfies $\psi ' = \theta$, and from this point onward the derivation of the equilibrium configuration is the same in both Euler Elastica and CP approaches. }

\section{Twisted ribbon derivation}\label{twisted_derivation}
We consider a ribbon geometry parameterized by coordinates
\[
(u,v,w)\in[-L/2,L/2]\times[-W/2,W/2]\times[-h/2,h/2],
\]
where \(L\), \(W\), and \(h\) denote the ribbon length, width, and thickness, respectively. 
We assume a thin-sheet limit with the scale separation \(h\ll W\ll L\). 
The reference configuration is taken to be flat, corresponding to a Euclidean reference metric
\(\bar{a}=\mathrm{Id}\) and vanishing reference curvature \(\bar{b}=0\).

We seek solutions of the FvK equations under these assumptions, and further assume
that there is no explicit dependence on the long coordinate \(u\). Accordingly,
the stress and curvature depend only on the transverse coordinate,
\(\sigma=\sigma(v)\) and \(b=b(v)\). We also impose traction-free boundary
conditions along the long edges of the ribbon 
$[\sigma\cdot n]_{v=\pm W/2}=0$ .
Under these assumptions, the most general form of the fields is
$f(u,v)=c_{1}u^{2}+c_{2}uv+\tilde{f}(v)$.
Substituting this ansatz into the FvK equations yields the general form of the Airy stress potential
\begin{equation}
\chi(u,v)=\frac{f_{0}}{2}u^{2}+f_{1}uv+\tilde{f}_{2}(v),
\qquad
\sigma=
\begin{pmatrix}
\tilde{f}_{2}''(v) & -f_{1}\\
-f_{1} & f_{0}
\end{pmatrix}.
\end{equation}
Imposing the traction-free boundary condition along the long edges of the ribbon
$[\sigma\cdot n]_{v=\pm W/2}=0$,
implies \(f_{0}=f_{1}=0\).

Under the
assumption of $u$–independence, the most general form of the CP is
\begin{equation}
\psi(u,v)=\frac{c_{0}}{2}u^{2}+c_{1}uv+\tilde{c}_{2}(v),
\qquad
b=
\begin{pmatrix}
c_{0} & c_{1}\\
c_{1} & \tilde{c}_{2}''(v)
\end{pmatrix}.
\end{equation}

Substituting this form into the FvK equations yields the reduced system
\begin{eqnarray}
\tilde{f}_2''''(v) &=& Y\left(c_1^{2} - c_0 \tilde{c}_2''(v)\right),
\label{eqdddd=detb}
\\
B \tilde{c}_2''''(v) &=& c_0 \tilde{f}_2''(v).
\label{eqdddd=sdd}
\end{eqnarray}

Choosing the boundary condition~\eqref{mnn} to be constant implies
\[
\tilde{c}_2(v)=\frac{1}{2}c_2 v^2 ,
\]
which in turn yields $\det(b)=\text{const}$. Equation~\eqref{eqdddd=detb}
then gives
\[
\tilde f_2(v)=\frac{1}{2}a_2 v^2+\frac{1}{6}a_3 v^3+\frac{1}{24}a_4 v^4 .
\]
Assuming symmetry about the ribbon midline sets $a_3=0$. Substituting this
into Eq.~\eqref{eqdddd=sdd} yields
\[
0=c_0\left(a_2+\frac{1}{2}a_4 v^2\right),
\]
from which we obtain $c_0=0$. The resulting solution therefore takes the form
\begin{eqnarray}
\chi(u,v) &=& \frac{a_2}{2}v^2+\frac{Y c^2_1}{24}v^4 , \notag\\
\psi(u,v) &=& c_{1}uv+\frac{c_2}{2}v^2 .
\end{eqnarray}

The configuration is controlled by three parameters determined by the
boundary conditions, where $Y$ is the Young modulus and $\nu$ is Poisson's ratio:
\begin{align}
\left[\sigma\cdot n\right]_{v=\pm w/2} &= 0,
&
\left[\sigma\cdot n\right]_{u=\pm L/2} &= a_2 + \tfrac{Y c^2_{1}}{2}\,v^{2},
\\[4pt]
\left[m\cdot n\cdot n\right]_{v=\pm w/2} 
&= \tfrac{h^{3}Y}{12(1+\nu)(1-\nu)}\,c_{2},
&
\left[m\cdot n\cdot n\right]_{u=\pm L/2} 
&= \tfrac{h^{3}Y\nu}{12(1+\nu)(1-\nu)}\,c_{2},
\\[4pt]
\left[\nabla(m)\cdot n\right]_{v=\pm w/2} &= 0,
&
\left[\nabla(m)\cdot n\right]_{u=\pm L/2} &= 0 .
\end{align}

Here $a_2$ controls the total end-to-end length of the ribbon, $c_1=\theta / L$ is the
twist per unit length (see Fig.~1(a) of the main text), and $c_2$ sets the curvature along the transverse
direction.

\begin{figure}
    \centering
    \includegraphics[width=0.5\linewidth,
        trim=80 160 190 210,        % L B R T in pt (adjust!)
        clip]{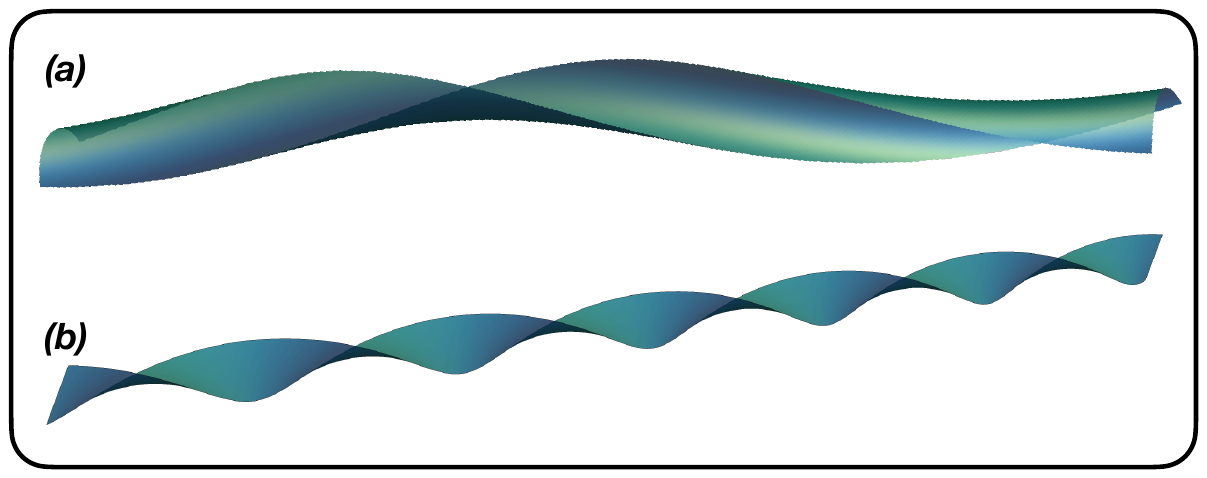}
   \caption{Resulting configurations of a twisted ribbon: 
Panel (a) shows a ribbon with dimensions $1\times10$ for $c_{1} = \tfrac{2\pi}{10}$, $a_2 = 0$, and $c_{2} = 3$; 
panel (b) corresponds to a ribbon with dimensions $1\times100$ with $c_{1} = \tfrac{6\pi}{100}$, $a_2 = 0$, and $c_{2} = 0$.}
    \label{fig:ribbons}
\end{figure}

\section{Bent sliced annulus}\label{open_disc}

\begin{figure}[h]
    \centering
    \includegraphics[
        width=0.15\linewidth,
        trim=80 150 80 80,
        clip
    ]{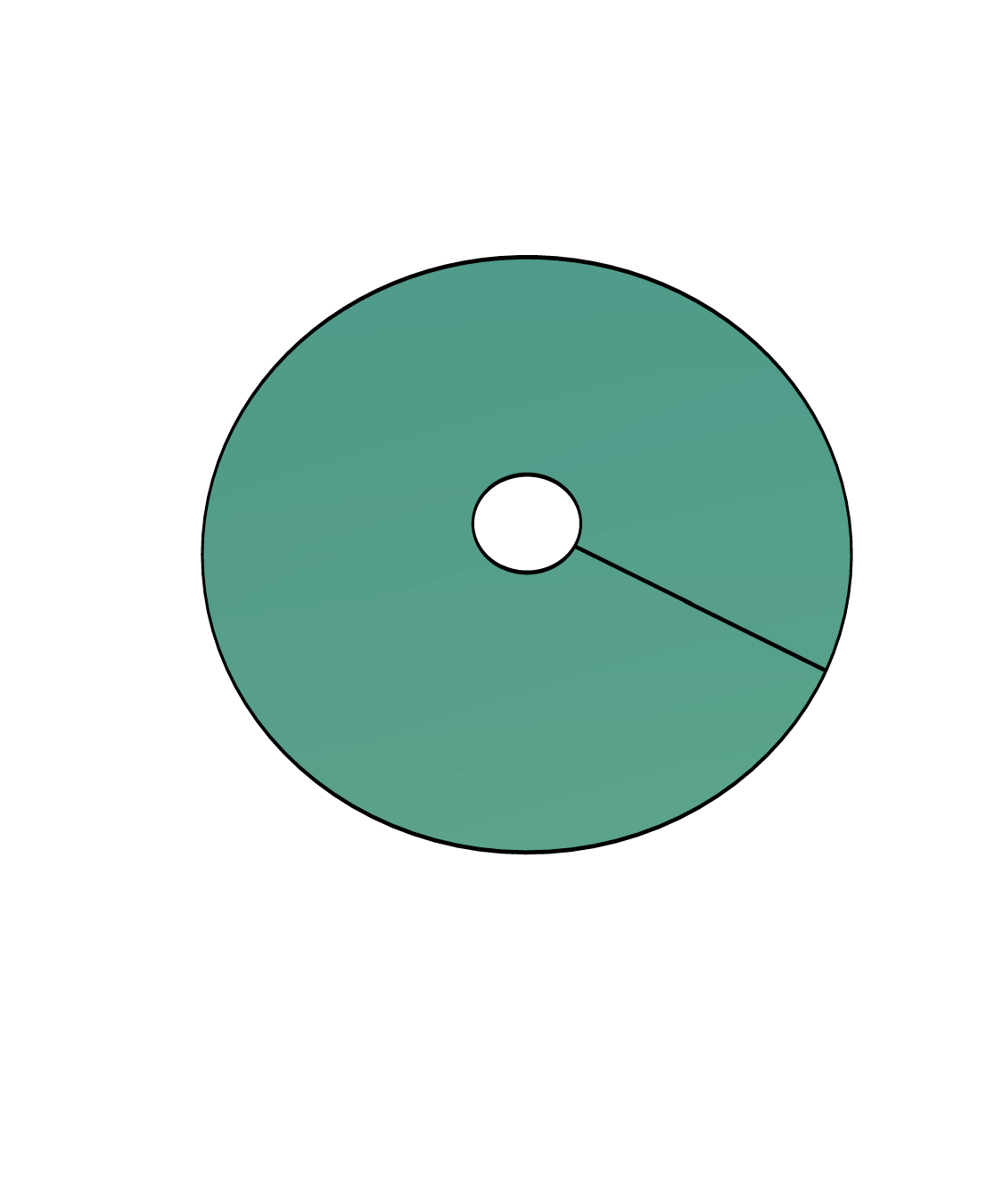}
    \ \ \ \ \ \
    \includegraphics[
        width=0.15\linewidth,
        trim=90 170 90 80,
        clip
    ]{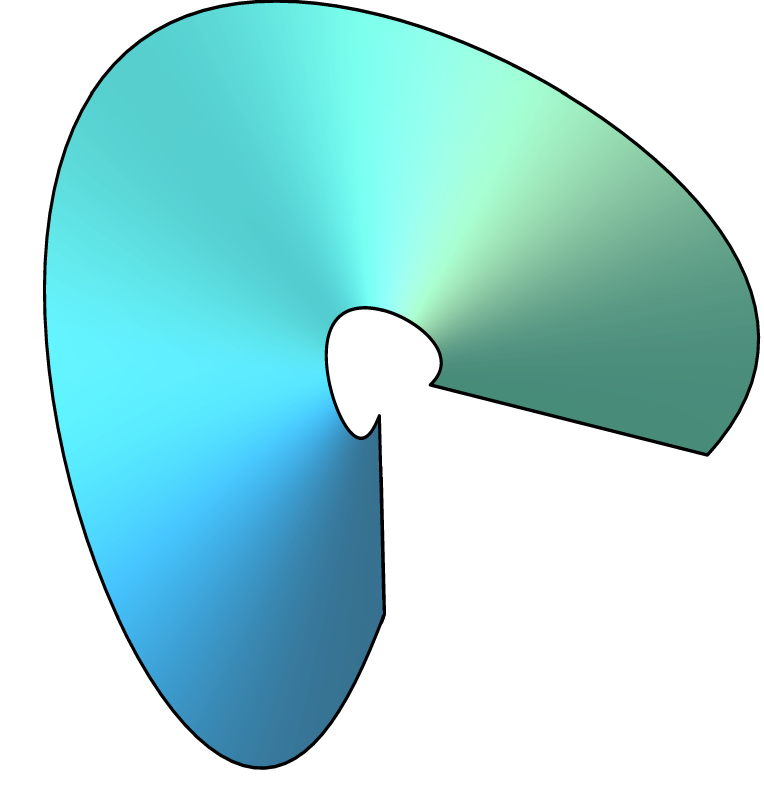}
    \ \ \ \ \ \
    \includegraphics[
        width=0.15\linewidth,
        trim=75 70 40 80,
        clip
    ]{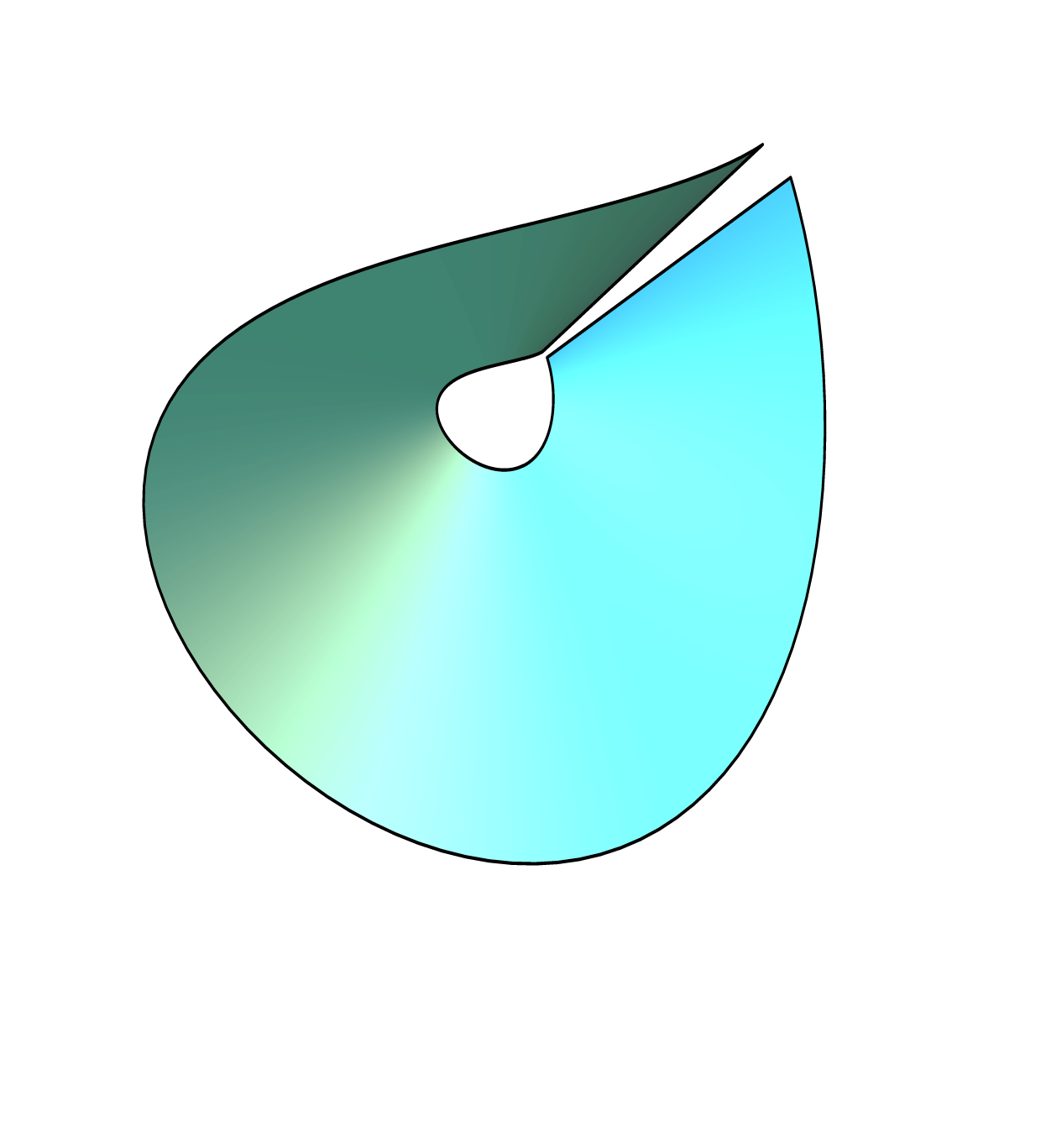}

    \caption{Opening an annulus with a radial cut. The resulting configuration
    is controlled by the opening angle and the relative tilt of the cut edges.}
    \label{fig:making_open_disc}
\end{figure}

We consider a flat annulus with a radial cut, which allows the sheet to be
opened by prescribing the angle between the two cut edges and their relative
tilt. Figure~\ref{fig:making_open_disc} illustrates several examples of such
configurations. The annulus is initially flat and has zero Gaussian
curvature, we search for configurations that remain stretching–free and are
therefore generated purely by bending.
 {This class of problems can also be formulated as Euler elastica on the unit sphere, as shown by Müller, Ben Amar, and Guven, "Conical defects
in growing sheets" PRL 2008. In the isometric limit, identifying $\partial_\theta\psi$ with the elastica angle field makes the two formulations equivalent and yields the same governing equation and solutions. The purpose of the present example, and of the CP framework more broadly, is to treat both extensible and inextensible sheets with geometrically complex configurations within a unified formulation based on the curvature potential.} 

A curvature potential that generates curvature only in the azimuthal direction
has the form $\psi(r,\theta)=r f(\theta)$. Solving the biharmonic equation for
$\psi$ yields the general solution
\begin{eqnarray}\label{eq:CP_annulus}
\psi(r,\theta)
= r\!\left(
c_{1}\,\theta\cos\theta
+ c_{2}\,\theta\sin\theta
+ c_{3}\cos\theta
+ c_{4}\sin\theta
\right).
\end{eqnarray}

For non isometric solutions, the corresponding SP $\chi$ that satisfies the equilibrium
equations has the form
\begin{eqnarray}
\chi(r,\theta)=
a_1 \cos(2\theta)+a_2 \sin(2\theta)
+ r\theta\left(b_{1}\cos\theta + b_{2}\sin\theta\right),
\end{eqnarray}
where additional terms that do not contribute to the stress have been omitted.

We focus on boundary conditions that admit stretching–free configurations,
namely isometric solutions with
$a=\bar{a}=
\begin{pmatrix}
1 & 0 \\
0 & r^{2}
\end{pmatrix}$.
For such solutions the SP vanishes ($\chi=0$), while the
curvature potential is given by Eq.~\eqref{eq:CP_annulus}.

The boundary conditions are obtained from the Weingarten reconstruction
(Appendix~\ref{b_t}). In this case the boundary force
\eqref{s_n} at $\theta=0,2\pi$ is
\begin{eqnarray}
\left[\Sigma\cdot n\right]_{\theta=0,2\pi}
=\left(0,\pm c_2^2\frac{Yh^3}{3r^4(\nu^2-1)}\right),
\end{eqnarray}
while the bending moment and its divergence
(\eqref{mnn},\eqref{dmn}) are
\begin{eqnarray}
\left[m\cdot n\cdot n\right]_{\theta=0,2\pi}
&=&\frac{c_2Yh^3}{6r^3(1-\nu^2)},\\
\left[\nabla(m)\cdot n\right]_{\theta=0,2\pi}
&=&\frac{\pm c_1 Yh^3}{6r^3(1-\nu^2)} .
\end{eqnarray}

To validate these analytic predictions we consider two representative
curvature potentials,
\begin{eqnarray}
\psi(r,\theta)=c ~r~\theta~\sin(\theta)\;, \quad \psi(r,\theta)=c ~r~\theta~\cos(\theta).
\end{eqnarray}
For each case the boundary conditions were obtained by numerically
integrating the Weingarten equations to reconstruct the configuration.
These boundary conditions were then imposed in finite–element simulations
of the full elastic problem.

\begin{figure}[h] \centering 
\includegraphics[width=\linewidth]{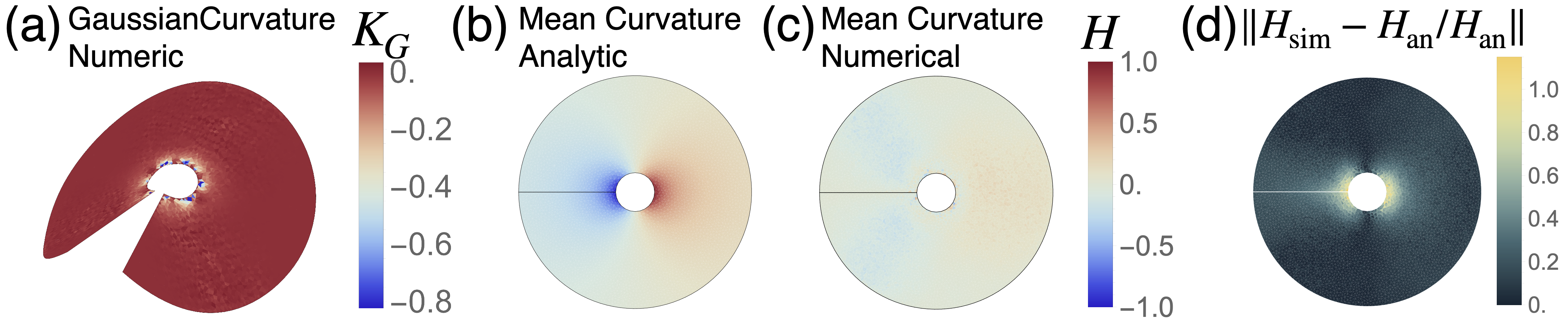} 
\caption{Finite–element simulation of an open annulus of dimensions $r_{in}=0.5, r_{out}=3$, thickness $h=0.04$, Young's modulus $Y=10^4$ and Poisson's ratio $\nu=0.3$, compared with the isometric analytic configuration with CP- $\psi(r,\theta)=0.4\, r\,\theta\sin\theta$. Shown are (a) Gaussian curvature, (b) analytic mean curvature, (c) simulated mean curvature, and (d) normalized deviation of the mean curvature. As in the previous case (Fig.~2(b) of the main text), the largest deviations occur near the inner boundary.} \label{fig:open_disc2} \end{figure}

Examples of the resulting configurations are shown in
Fig.~2(b) of the main text and \figref{fig:open_disc2}. In
\figref{fig:open_disc2} we compare the Gaussian and mean curvature fields
obtained from the simulations with the analytic predictions derived from the
curvature potential with $\chi=0$. The configuration is colored by Gaussian
curvature (which vanishes in the isometric limit), together with maps of the
simulated mean curvature, the analytic mean curvature, and their deviation.

\nocite{kovsmrlj2013mechanical}

%\bibliography{references}
%\end{document}

\end{document}